# Ultra-Low Power Neuromorphic Obstacle Detection Using a Two-Dimensional Materials-Based Subthreshold Transistor

Kartikey Thakar[1], Bipin Rajendran[2], and Saurabh Lodha[1,*]
[1]Department of Electrical Engineering, Indian Institute of Technology Bombay, Mumbai, India.
[2]Department of Engineering, King's College London, Strand, London, United Kingdom.
*slodha@ee.iitb.ac.in

## Abstract
Accurate, timely and selective detection of moving obstacles is crucial for reliable collision avoidance in autonomous robots. The area- and energy-inefficiency of CMOS-based spiking neurons for obstacle detection can be addressed through the reconfigurable, tunable and low-power operation capabilities of emerging two-dimensional (2D) materials-based devices. We present an ultra-low power spiking neuron built using an electrostatically tuned dual-gate transistor with an ultra-thin and generic 2D material channel. The 2D subthreshold transistor (2D-ST) is carefully designed to operate under low-current subthreshold regime. Carrier transport has been modelled via over-the-barrier thermionic and Fowler-Nordheim contact barrier tunnelling currents over a wide range of gate and drain biases. Simulation of a neuron circuit designed using the 2D-ST with 45 nm CMOS technology components shows high energy efficiency of ~3.5 pJ/spike and biomimetic class-I as well as oscillatory spiking. It also demonstrates complex neuronal behaviors such as spike-frequency adaptation and post-inhibitory rebound that are crucial for dynamic visual systems. Lobula giant movement detector (LGMD) is a collision-detecting biological neuron found in locusts. Our neuron circuit can generate LGMD-like spiking behavior and detect obstacles at an energy cost of <100 pJ. Further, it can be reconfigured to distinguish between looming and receding objects with high selectivity.

# Introduction

Brain-inspired neuromorphic computational paradigm offers the promise to revolutionize embedded energy-efficient decision-making systems by mimicking the event-triggered learning and inference aspects of the human brain. The hardware implementations of this paradigm can also mitigate the bottleneck in throughput via in-memory computing enabled by novel nanoscale devices[1-4]. Simultaneous advances in both electronic hardware and software algorithms in the recent years with a focus to move towards low-power, low-area and time-efficient compute applications have fueled the progress in neuromorphic engineering. Spiking neural networks (SNNs) take inspiration from the brain for their architecture and consist of individual spiking neurons (smallest 'computation' element) interconnected with each other through synapses (functional connection between the neurons). SNNs have been shown to perform complex computations with flexibility and robustness using minimal resources[5,6].

Autonomous robots and vehicles have been seeing significant growth and adoption by industry for manufacturing and transportation. Autonomous movement in unknown terrains requires efficient object tracking and timely obstacle detection capability. Obstacle detection and selective response for approaching and receding objects are highly useful for real-time path planning. Existing solutions based on VLSI vision systems with complex algorithms offer performance at the cost of area- and energy-inefficiency[7-10]. Many biological species have collision detecting neurons in their visual pathway to differentiate and detect threats based on the trajectory of incoming objects in their fields of view[11-15]. For example, locusts have specialized looming-sensitive neurons, called Lobula giant movement detectors (LGMDs), in their visual neural pathway that are capable of detecting a possible collision with approaching objects within a few milliseconds. LGMD neurons encode the dynamics of incoming objects in their firing rate based on the size and velocity of the object[12,16,17]. A fascinating detail about locust neurobiology is that they have only two such neurons (one per eye) to help them avoid collisions with precision and efficiency. Another important behavior of LGMD is its selectivity for looming collision threats while receding objects do not elicit a noticeable spike response. Such selectivity is crucial from a safety viewpoint especially when approaching objects are in direct collision path. Hence, LGMDs present an excellent example of a spiking neuron that efficiently implements the computational dynamics involved in a complex cognitive task with minimal resources.

In this work, we demonstrate a low-power neuron that closely matches the essential computational features of the LGMD neuron using a subthreshold transistor based on two-dimensional (2D) materials. 2D materials are good candidates for novel device architectures with tunable and low-power CMOS-compatible operation[18-25]. We engineer a dual-gate 2D transistor that can generate a low-current bell-shaped current-voltage (I-V) transfer curve under subthreshold operation which is tunable via electrostatic control from the two gate-bias signals. The bell-shaped I-V is useful in mimicking fast activation and self-inactivation of Na-channels in biological neurons as described in the Hodgkin-Huxley (HH) model[26,27]. Transport in the 2D subthreshold transistor (2D-ST) is shown to be governed by two separate physical phenomena, over-the-barrier thermionic current and Fowler-Nordheim (FN) tunneling current – through physics-based modelling of the bell-shaped I-V curves. The device design and operation are engineered to give a two-fold advantage over existing literature: (i) low-current operation to achieve high energy efficiency, and (ii) facile fabrication with low complexity by using a single material channel without compromising tuning capability via electrostatic control.

Next, we design a low-power circuit that incorporates the behavioral model of our 2D-ST device with 45 nm CMOS to show biomimetic class-I regular spiking behavior with low power (~3.5 pJ/spike) operation. Simulations show that the circuit can be adapted to function as an oscillatory neuron by adjusting the conductance of the leakage path, and can also be modified to exhibit spike frequency adaptation (SFA) and post-inhibitory rebound (PIR) spikes. Both SFA and PIR are important functionalities demonstrated by biological LGMD neurons[28,29]. We finally demonstrate that the neuron circuit based on our 2D-ST can mimic the behavioral features of an LGMD neuron under varying test conditions by modelling the synaptic current to for looming (approaching) as well as receding objects. We show that the circuit is consistently able to detect looming objects prior to collision with a small number of spikes (<30) and energy dissipation of <100 pJ. Furthermore, the artificial neuron circuit is also able to effectively match the LGMD functionality of differentiating between looming and receding objects. This inherent selectivity to approaching objects helps in prioritizing the system response to impeding collisions with obstacles in the direct path. Going another step forward, we also show how the LGMD circuit can be easily reconfigured to give either looming or receding object-selective spike response. Such flexibility adds a degree of freedom in real-time multi-object tracking system design with distinct responses for the speed and direction of moving objects in the field of view.

This work demonstrates a spiking neuron circuit with biomimetic LGMD functionality with low-power spikes for energy-efficient obstacle detection applications. This would allow seamless inclusion of such circuits with always-on, low power SNN-based systems that operate using event-driven signal processing and computational algorithms. Additionally, the 2D-ST neuron also improves the power dissipation by several orders of magnitude over existing literature on biomimetic spiking neurons ($\times 10^5$) and LGMD applications ($\times 10^2$) with novel material systems[30-32]. Energy efficiency can be improved further with the use of scaled transistor technologies[33,34].

# Results and Discussion

## 2D subthreshold transistor

The design of the 2D transistor is driven by the requirement of a low-current bell-shaped current-voltage transfer curve ($I_D$-$V_G$) that mimics the voltage-dependent conductance of biological Na channels so as to achieve bio-realistic spike frequencies with low energy dissipation. The bell-shaped curve with the negative and positive transconductance regions is used to implement positive feedback for spike initiation and self-inactivation for spike reset. This requirement is engineered in our device with simultaneous electrostatic control of (i) the 2D channel barrier to tune the conductance of the device, and (ii) the Schottky barriers to tune the FN tunneling-driven current injection. This dual electrostatic control is implemented in our device with a channel gate ($G_C$) and a source/drain Schottky barrier gate ($G_B$). Low device current is obtained via subthreshold operation under an appropriate biasing scheme. **Figure 1**a shows the fabrication steps for a 2D-ST. The devices were fabricated on $Si/SiO_2$ substrates, starting with the patterning of the $G_C$ and $G_B$ back-gates. Next, mechanically exfoliated hBN was transferred over the patterned gates as bottom-gate dielectric followed by dry transfer of the 2D semiconducting material. Three different 2D semiconductors viz. $MoS_2$, $WSe_2$ and $WS_2$ were tried out for this work showing that a bell-shaped I-V curve can be obtained irrespective of the channel material and its propensity for being n- or p-type. Lastly, source/drain metal electrodes were fabricated with an overlap with $G_B$ for efficient Schottky barrier control. Although mechanically exfoliated flakes were chosen for this study, the same architecture can be used for scalable manufacturing with large-area growth of the 2D dielectric and semiconducting materials.

Optical image of an as-fabricated 2D-ST with an $MoS_2$ channel is shown in Figure 1b. Thicknesses of the hBN and $MoS_2$ flakes were determined to be ~35 nm and ~3.5 nm, respectively, using atomic force microscopy (AFM) measurements (Supplementary Material (SM) Figure S1). Device current for varying gate bias ($V_{GC}$, $V_{GB}$) values for the $MoS_2$ device is plotted in Figure 1c. n-type $MoS_2$ gives negligible off current (~5 pA) under negative bias ($V_{GC} = V_{GB} = -4$ V) values and reaches high on-current (~600 nA) for positive bias ($V_{GC} = V_{GB} = 1$ V) values. Current trace extracted from Figure 1c for the condition $V_{GB} = V_{GC}$ gives conventional n-FET behavior (red curve on the top right) as expected. However, a low-current (~15 nA) bell-shaped curve (blue curve on the top left) can be achieved by taking the current trace along the opposite direction ($V_{GB} \equiv -V_{GC}$) which forces the device to operate in subthreshold regime.

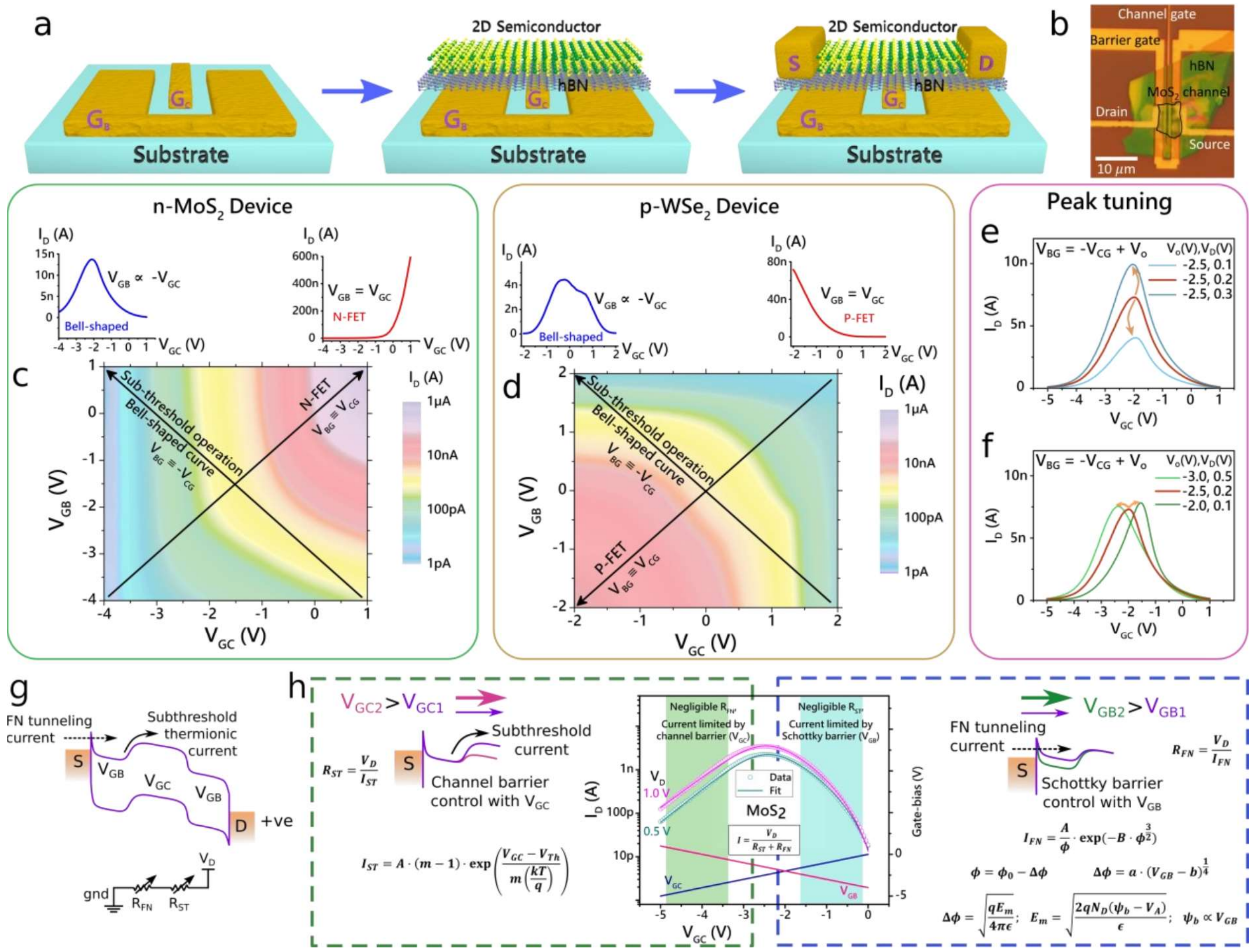

**Figure 1. 2D-ST transistor design and operation. (a)** Schematics illustrating typical fabrication flow for the 2D subthreshold transistor (2D-ST) with multi-gate FET structure, hBN gate dielectric and *n*- or *p*-type 2D semiconductor channel. The two gates – channel gate ($G_C$) and barrier gate ($G_B$) – control the 2D channel barrier and the Schottky barriers at the source/drain contacts. **(b)** Optical image of an as-fabricated 2D-ST with a few-layer *n*-type $MoS_2$ channel. **(c)** Colormap showing device current as a function of the channel and barrier gate biases ($V_{GC}$, $V_{GB}$). As expected, the device shows *n*-type FET behavior. The line scan along $V_{GC} = V_{GB}$ direction from the current map results in a conventional n-FET transfer curve. Whereas, taking a line scan along the opposite direction ($V_{GC} \equiv -V_{GB}$) gives a bell-shaped curve under subthreshold operation. Subthreshold regime is targeted for low current values. Bell-shaped curve is useful to mimic the sodium (Na) channel behavior in a biomimetic spiking neuron circuit. **(d)** This technique to obtain a bell-shaped curve is applicable irrespective of the 2D channel material and dominant conduction polarity (*n*- or *p*-type). Results similar to (c) are shown with a *p*-type $WSe_2$ channel showing a bell-shaped curve for $V_{GC} = -V_{GB}$. Additionally, the bell-shaped curve is tunable by electrostatic control *via* $G_C$ and $G_B$. Plots showing tuning of peak current **(e)** value and **(f)** position with appropriate biasing with the *n*-$MoS_2$ device. **(g)** Current through the device is controlled via Fowler-Nordheim tunneling current ($V_{GB}$) and over-the-barrier thermionic current ($V_{GC}$). The two current components can be modelled by two variable resistors ($R_{FN}$ and $R_{ST}$) in series. **(h)** $R_{FN}$ (blue) and $R_{ST}$ (green) can be estimated through region-wise fits of the bell-shaped I-V curve using analytical equations. Center plot shows fitting of the complete I-V curve by combining the effects of both the resistors.

The proposed biasing scheme is applicable irrespective of the channel material and its charge polarity (n- or p-type). Figure 1d shows results similar to Figure 1c with a p-type $WSe_2$ channel. Current trace along the $V_{GB} = V_{GC}$ direction gives a conventional p-FET response (red curve) whereas a trace along the $V_{GB} = -V_{GC}$ direction results in a bell-shaped curve (blue curve) with low current levels (~4 nA). Similar results were obtained with n-type $WS_2$ as well (SM Figure S2). Moreover, due to the degree of freedom provided by electrostatic gating, these bell-shaped current peaks can be further tuned for desired peak current value or position by setting an appropriate offset voltage $V_o$ ($V_{GB} = -V_{GC} + V_o$) and $V_D$ (Figures 1e,f). Such tunability could be leveraged in device operation to achieve tunable Na-channel current in the biomimetic neuron design.

Band diagram in Figure 1g illustrates how the two gate voltages $V_{GC}$ and $V_{GB}$ control the thermionic current at the source barrier and FN tunneling at the contacts, respectively. An equivalent model with variable resistors $R_{ST}$ and $R_{FN}$ corresponding to the two transport mechanisms is shown, where the total resistance of the device is $R = R_{ST} + R_{FN}$. In the subthreshold regime, the device current is limited by the source-to-channel barrier (governed by $V_{GC}$) when carrier injection is not limited (large $V_{GB}$) on the left of the bell-shaped curve. The operation gradually switches to carrier injection limit (negative $V_{GB}$) with no source-to-channel barrier (large $V_{GC}$) towards the right of the curve as shown in the I-V curve ($MoS_2$ device) in Figure 1h. A physics-based region-wise fitting was performed to elucidate the transport regimes limited by over-the-barrier transport ($V_{GC}$) on the left (green) and FN tunneling ($V_{GB}$) on the right (blue). The set of equations used to model each transport regime along with illustrative partial energy band diagrams are shown on either side of the I-V curve in Figure 1h. Here, we can achieve low-current bell-shaped I-V by leveraging the device architecture to engineer its operation via simultaneous control of the two transport mechanisms. The fitting of the bell-shaped I-V curves is discussed in detail in SM Section S3 along with the correlation of extracted parameters with the physical device dimensions. We note that with the same set of model parameters, we are able to obtain excellent agreement between the model and experimental data over a wide range of operating conditions with high confidence, confirming that the two-regime model can accurately explain the underlying physics behind the bell-shaped operating curve. We also note that the underlying physics behind our devices is hence different from previously reported Gaussian transistors[32,35].

## Biomimetic neuron

The axon hillock neuronal circuit incorporating a Verilog-A model of the 2D-ST transistor along with components from the 45-nm CMOS general process design kit (gpdk) is shown in **Figure 2**a. Here, $V_M$ denotes the membrane potential at the axon hillock which collects the aggregated synaptic current from the pre-neurons (represented by a current source $I_{syn}$ in the simulations) and also serves as the output node for the post-neurons. $C_M$ is the membrane capacitance for the cell. Na-channel is mimicked by the 2D-ST as the pull-up transistor $T_{Na}$ whose current is governed by the two gate inputs $V_{GC}$ and $V_{GB}$. The gate biases of $T_{Na}$ are driven by two inverter pairs (I1: $MP1_{Na}$-$MN1_{Na}$, and I2: $MP2_{Na}$-$MN2_{Na}$) that enable generation of the polarizing Na-channel current ($I_{Na}$) based on the membrane potential $V_M$, while ensuring that $V_{GC}$ and $V_{GB}$ move in opposite directions (i.e., $V_{BG} \sim -V_{GC}$) to generate the required bell-shaped I-V response. Na-activated K-channel current ($I_K$) is implemented by the transistor $T_K$ driven by a delay element ($MN_K$-$C_K$) to mimic a delayed response activated mainly by the large $I_{Na}$ flowing in the circuit after Na-channel activation. A leakage path with high resistance is provided with the transistor $T_L$ in OFF condition. Conductance of $T_L$ can be varied to tune the threshold input current ($I_T$) required to elicit a spike response.

To understand the circuit operation, consider the initial condition where all capacitors are discharged and the membrane is at rest potential ($V_M = V_{rest}$). (1) Input synaptic current ($I_{syn}$) from pre-neurons starts charging $C_M$ with a $V_M$-dependent leakage current $I_L$ flowing through $T_L$. (2) Once $V_M$ reaches the turn-on threshold of the inverter pair I1, the gate bias $V_{GC}$ increases while the gate bias $V_{GB}$ starts to decrease through the action of the inverter I2; causing a large (compared to $I_{syn}$) depolarizing current $I_{Na}$ to flow from $T_{Na}$. (3) This excess current $I_{Na}$ starts charging $C_K$ rapidly to activate the delayed response from $T_K$. (4) Once $T_K$ is activated, it works as a sink to discharge $C_M$ (equivalent to a polarizing current) which brings $V_M$ down rapidly and switches off $I_{Na}$ as well in the process. These steps complete one cycle for action potential (spike) generation. A single spike generated for a 10 µs/500 pA $I_{syn}$ pulse is plotted in Figure 2b along with $I_{Na}$ and $I_K$ with inward current shown as positive. This circuit behaves as a regular spiking neuron as observed from its response to a continuous 500 pA $I_{syn}$ stimulus (Figure 2c). Its equivalence with the HH model is discussed in SM Figure S4. Furthermore, the circuit behavior as a leaky-integrate-and-fire (LIF) neuron is probed with variable frequency and pulse width of the $I_{syn}$ pulses as detailed in SM Figure S5.

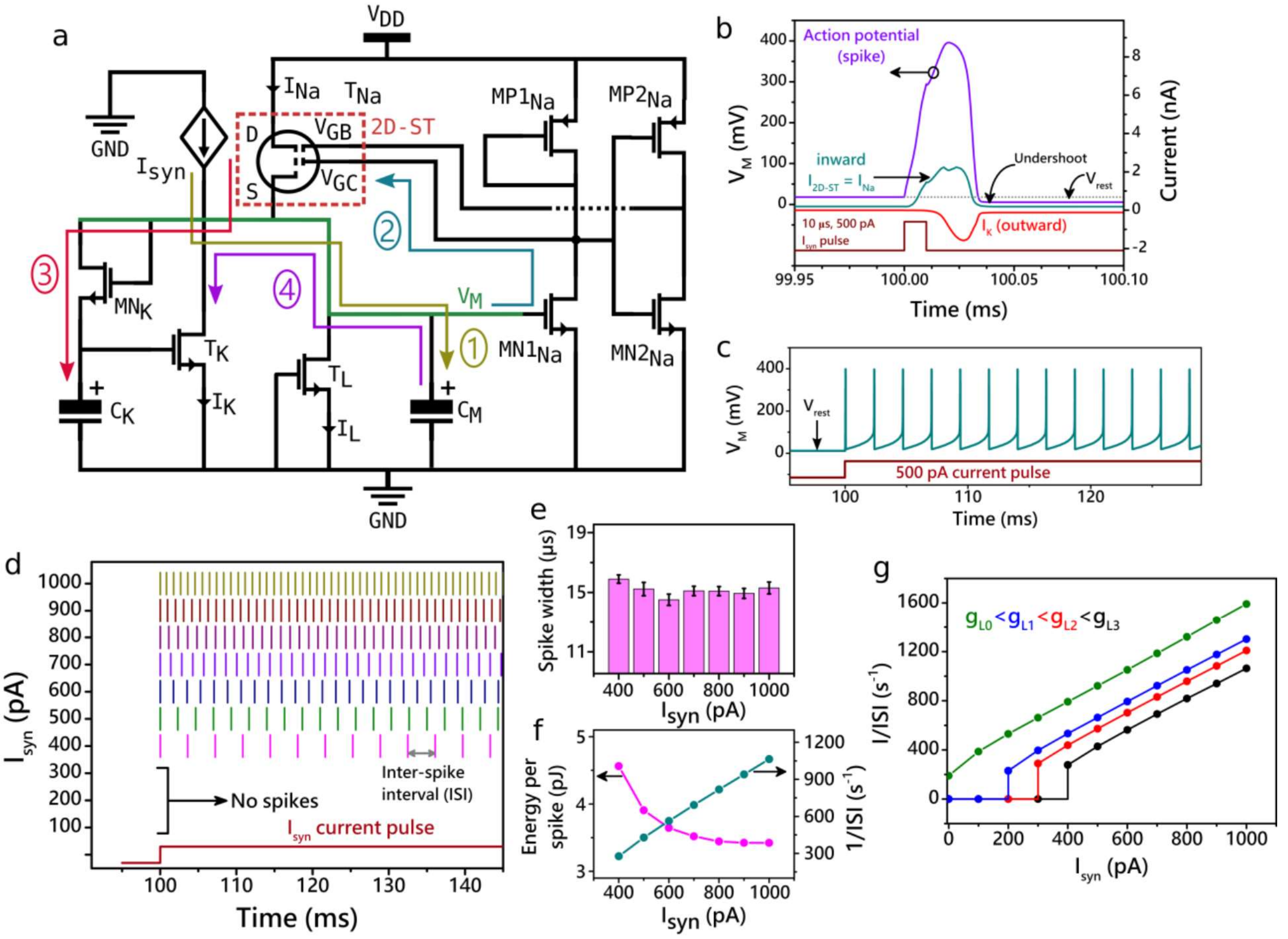

**Figure 2. Biomimetic neuron with 2D subthreshold transistor. (a)** Schematic of the simulated circuit showing 2D-ST to mimic sodium (Na) channel current. The two inverters ($MN1_{Na}$-$MP1_{Na}$ and $MN2_{Na}$ - $MP2_{Na}$) are used to drive the two gates ($V_{GC}$ and $V_{GB}$) of the 2D-ST. $T_K$ with a delay element ($MN_K$-$C_K$) provides a delayed potassium (K) channel current. $C_M$ is the membrane capacitor. $T_L$ represents the leakage path. $V_{DD}$ is the supply and $I_{syn}$ is used to supply input synaptic current. All transistors (MOSFETs) and capacitors (MOSCAPs) except for the 2D-ST are taken from 45 nm pdk. The node $V_M$ is where the action potential, or spike, is recorded. **(b)** Simulation result for a 10 µs, 500 pA $I_{syn}$ pulse shows generation of a single spike. Na and K channel currents are also plotted for reference with inward current (with respect to $V_M$) shown as positive. An undershoot with respect to the rest potential ($V_{rest}$) is also observed as shown. **(c)** Circuit response showing regular spiking behavior for a continuous 500 pA input current. **(d)** Raster plot showing spike activity as a function of input current magnitude. Low current values up to 300 pA do not elicit spikes indicating a threshold input current ($I_T$) to generate spikes. Increasing current value from 400 pA to 1000 pA gives consistent spikes with monotonically increasing spike density. The interval between two consecutive spikes, or inter-spike interval (ISI), is used to record the instantaneous spike rate or frequency. **(e)** Spike width measured at $V_M$ = 300 mV shows negligible variation with $I_{syn}$ magnitude. **(f)** Energy per spike varies between 4.5-3.5 pJ with $I_{syn}$, whereas spike rate (or $ISI^{-1}$) increases linearly with $I_{syn}$. **(g)** $I_T$ can be adjusted by changing the conductance of $T_L$ where reducing the conductance leads to reduced $I_T$ and increased spike rate. Data shown in (b-f) corresponds to $g_{L3}$. The case corresponding to $g_{L0}$ shows non-zero spikes at $I_{syn} = 0$ suggesting that the circuit can also be used as an oscillatory neuron.

Reduction in $I_{Na}$ with $V_M$ (self-inactivation) is achieved in this circuit, thanks to the bell-shaped I-V characteristics of the 2D-ST. This is crucial in reducing the excess current through $C_K$ with $V_M$ and allowing it to discharge through $T_K$. It is important to note that once the Na-channel current is activated, the circuit goes into a self-sustaining loop to complete the entire cycle of spike generation irrespective of the $I_{syn}$. This is a notable characteristic behavior observed in biological neurons as well. Additionally, once $T_K$ is activated, it can drive $V_M$ below $V_{rest}$ (hyperpolarization) to give an undershoot with respect to $V_{rest}$ until its gate voltage, decided by $MN_K$-$C_K$, falls back towards zero (ground). Threshold voltage of the 2D-ST is crucial to set its region of operation that is compatible with the rest of the transistors in the circuit. The fabricated 2D-ST devices operate in the 2 V range as the effective oxide thickness (EOT) of these devices is ~40 nm compared to the 0-1 V operating regime of the nominal 45 nm CMOS transistors whose EOT is ~2.5 nm. Hence, to account for the EOT and threshold voltage difference during circuit evaluation, behavioral characteristics of p-$WSe_2$ 2D-ST device were used after scaling the gate bias as $V' = V/4 + 0.5$ to map the range of ±2 V to the nominal range of 0-1 V for the 45 nm pdk. We have also designed and simulated a similar circuit that uses R-C components instead of the CMOS components operating at a larger $V_{DD} = 2$ V to accommodate for the wider gate-bias range (±2 V) of the p-WSe2 device without scaling (SM Figure S6) to demonstrate biomimetic LIF neuron behavior with energy per spike of ~500 pJ.

Circuit simulations were repeated for $I_{syn}$ magnitudes ranging from 100 pA to 1000 pA and their results are summarized in Figure 2d. Each bar (|) in Figure 2d represents a spike. No spike is generated for $I_{syn}$ = 100-300 pA indicating an $I_T > 300$ pA. Class-I regular spike response[36] was observed for $I_{syn} \geq$ 400 pA with the spike rate increasing with $I_{syn}$ magnitude. Instantaneous spike rate can be represented by the duration between two consecutive spikes, denoted as inter-spike interval (ISI), with a smaller ISI meaning a larger spike rate. Spike width measured at $V_M = 300$ mV corresponding to the circuit response for a range of $I_{syn}$ values showed little variation as shown in Figure 2e. Energy per spike (EPS) is calculated as the energy dissipated by all the circuit components during one complete spike (or the duration between two consecutive spikes). Figure 2f shows the evolution of EPS and $ISI^{-1}$ with $I_{syn}$. An EPS value as low as ~3.5 pJ makes this circuit useful for low-power spiking neuron applications. An exercise similar to Figure 2d was repeated for different leakage conductance ($g_L$) values by changing the W/L ratio of the transistor $T_L$. Smaller value of $g_L$ can results in smaller $I_T$ as seen in Figure 2g, and can also achieve a spiking response for $I_{syn} = 0$. Such a circuit can be used as an oscillatory neuron which spikes periodically at a constant rate without any external synaptic current input. This configuration can be useful in the implementation of thalamocortical networks[37,38]. Details of each of the components of the circuit in Figure 2a are listed in SM Table S7-1.

## Spike frequency adaptation (SFA) and post-inhibitory rebound (PIR)

In addition to regular spiking behavior, biological neurons show a rich repertoire of spiking patterns[36]. These different spiking behaviors serve different functional requirements of neural networks. SFA and PIR are two such features that can enable complex neural computations at the single cell-level. SFA in neurons has been shown to play an important role in several tasks involving cognition, inference, and memory[39-41]. Approximately 40% of neocortical excitatory neurons in the human brain show SFA functionality. SFA implies that for a constant input synaptic current, spike rate of the neuron decreases gradually with time (as opposed to a constant spike rate corresponding to a constant current input in regular spiking) with time constants determined by the underlying biochemical mechanisms. Several ionic channels have been shown to contribute to spike rate adaptation. $Ca^{2+}$-activated K channels (polarizing current) are one of the important ionic channels that enable SFA functionality[42,43].

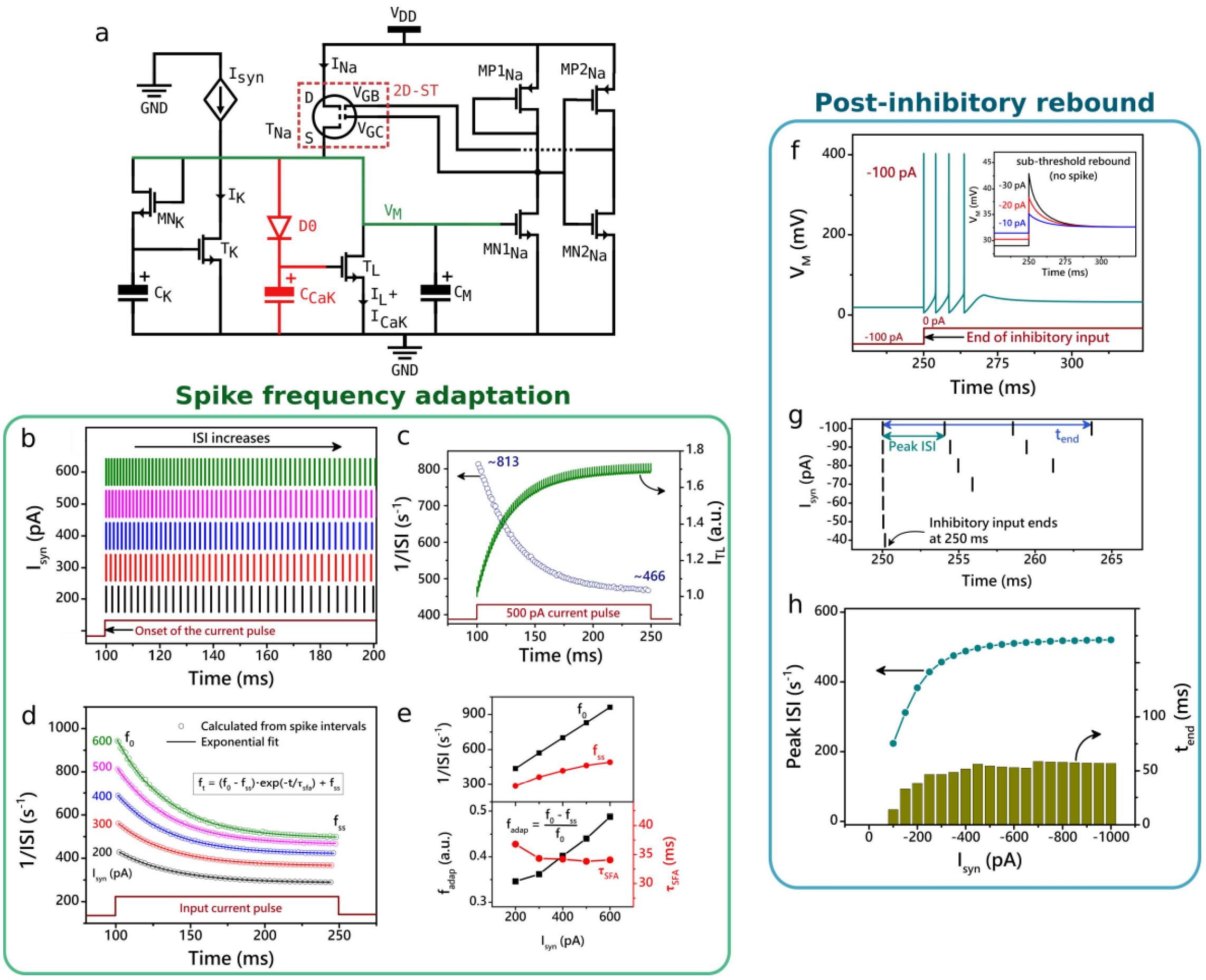

**Figure 3. Spike frequency adaptation and post-inhibitory rebound. (a)** Schematic of the modified circuit to demonstrate spike frequency adaptation by mimicking calcium-activated potassium channel ($K_{Ca}$). A diode D0 in series with a capacitor $C_{CaK}$ (marked in red) lead to a non-linear, delayed rise in hyperpolarizing current with help of the transistor $T_L$. **(b)** Raster plot showing spike frequency adaptation by simulating the circuit in (a) for a range of amplitudes (200-600 pA) of the $I_{syn}$ pulses. ISI increases with time, or in other words, spike frequency reduces. **(c)** Instantaneous ISI and current through $T_L$ ($I_{TL}$) as a function of time for 500 pA $I_{syn}$ pulse. $I_{TL}$ emulates both leakage current ($I_L$) and $K_{Ca}$ current ($I_{CaK}$). $ISI^{-1}$ decreases and saturates to a lower value in response to an increase in $I_{TL}$. **(d)** Instantaneous $ISI^{-1}$ as a function of time similar to (b) for the data in (a). The data is fitted with an exponential decay function to extract initial spike frequency ($f_0$), steady-state spike frequency ($f_{ss}$) and the decay time constant ($\tau_{SFA}$). Adaptation factor ($f_{adap}$) is calculated as $(f_0-f_{ss})/f_0$. **(e)** Plots showing parameters extracted from (d) with $I_{syn}$. As expected, $f_0$ increases with $I_{syn}$ with lower $f_{ss}$ values. $\tau_{SFA}$ does not vary with $I_{syn}$ as expected, and $f_{adap}$ increases with $I_{syn}$ suggesting a limit on the spike rate because of $I_{CaK}$. **(f)** Functionally, the same components (D0-$C_{CaK}$) also help elicit a rebound signal after an inhibitory (negative) current input. $V_M$ plot after the end of a -100 pA $I_{syn}$ pulse shows spike generation. Inset shows increasing post-inhibitory rebound in $V_M$ for lower $I_{syn}$ magnitudes (10-30 pA) which fail to generate spikes. **(g)** Raster plot showing spike timing for inhibitory $I_{syn}$ magnitudes. The number of spikes and $ISI^{-1}$ increase with increasing negative $I_{syn}$. The time of the last spike occurrence with respect to end of inhibitory input is recorded as $t_{end}$. **(h)** Peak $ISI^{-1}$ and $\tau_{end}$ increase reaching saturation with more negative $I_{syn}$. Saturation in $t_{end}$ and its similarity to $\tau_{SFA}$ reaffirms the role of D0-$C_{CaK}$ in evoking post-inhibitory rebound.

SFA can be achieved using the circuit in Figure 2a with the minimal modification of adding a delay element (D0-$C_{CaK}$) to drive the leakage path $T_L$ as shown in **Figure 3**a. These changes make $T_L$ perform two tasks simultaneously – (i) provide a constant leakage path (equivalent to when $C_{CaK}$ is completely discharged), and, (ii) provide a time-dependent delayed K channel current $I_{CaK}$ (different from Na-activated K channel current $I_K$) controlled by D0-$C_{CaK}$. Hence, current through $T_L$ can be expressed as $I_{TL} = I_L + I_{CaK}$. Details of each of the components of the circuit in Figure 3a are listed in SM Table S7-2. Figure 3b shows a raster plot showing spike activity (for the circuit in Figure 3a) for a range of $I_{syn}$ amplitudes. The reduction in spike rate (or increase in ISI) with time for no change in $I_{syn}$ indicates SFA behavior.

To elucidate the role of $I_{CaK}$ in SFA, instantaneous $ISI^{-1}$ is plotted for $I_{syn}$ = 500 pA as a function of time along with $I_{TL} = I_L + I_{CaK}$ in Figure 3c. Since $I_L$ is very small compared to $I_{CaK}$, and does not vary with time, we can assume $I_{TL} \approx I_{CaK}$. Decrease in $ISI^{-1}$ shows one-to-one correspondence with increase in $I_{TL}$. Temporal $ISI^{-1}$ trends for all $I_{syn}$ magnitudes (200-600 pA) were fitted with an exponential decay function $f(t) = (f_0 - f_{ss}) \cdot \exp\left(-t/\tau_{SFA}\right) + f_{ss}$ (Figure 3d). The extracted parameters - initial spike frequency ($f_0$), steady-state spike frequency ($f_{ss}$), and the decay time constant ($\tau_{SFA}$) - are plotted in Figure 3e along with the frequency adaptation factor calculated as $f_{adap} = (f_0 - f_{ss})/f_0$. Both $f_0$ and $f_{ss}$ increase with $I_{syn}$ (top plot) with $f_{ss}$ showing saturation, resulting in an increasing $f_{adap}$ as shown in the bottom plot. Note that $\tau_{SFA}$ remains largely unchanged and independent of $I_{syn}$ as expected, since it is determined by the time constant of the D0-$C_{CaK}$ pair.

In addition to SFA, this same circuit was able to elicit PIR behavior as well in response to inhibitory (negative) $I_{syn}$. PIR refers to an overshoot in the membrane potential of a neuron after an inhibitory current input ends. One or multiple spikes may be generated by the neuron based on the magnitude of the rebound potential. PIR is an important neuronal behavior useful for complex tasks in cognition and short-term memory retention[44-47]. Figure 3f shows the $V_M$ response after the end of an $I_{syn}$ = -100 pA pulse at 250 ms. Multiple spikes were recorded with increasing ISI before $V_M$ settled back down to $V_{rest}$. Inset shows PIR for small negative $I_{syn}$ values (-10 to -30 pA) which fail to generate a spike. Gradual relaxation in $V_M$ with time indicates that the rebound signal is triggered mainly by the excess negative charge stored in the capacitor $C_{CaK}$ which turns off $T_L$. Figure 3g shows a raster plot of spikes for a range of $I_{syn}$ values (-40 to -100 pA) suggesting that both the spike count as well as the spike rate increase with increasingly negative $I_{syn}$. The peak $ISI^{-1}$ is calculated from the ISI for the first two spikes and duration from the end of the inhibitory input (here, 250 ms) to the occurrence of the last spike is recorded as $t_{end}$. These parameters (peak $ISI^{-1}$ and $t_{end}$) are plotted in Figure 3h. Both the peak $ISI^{-1}$ and the $t_{end}$ saturate with $I_{syn}$. Similarity of $t_{end}$ and $\tau_{SFA}$ values indicates that the D0-$C_{CaK}$ pair plays a key role in eliciting both SFA and PIR response. In addition, the normalized decay profiles for the spikes with time (shown in SM Figure S8) were identical for all $I_{syn}$ values further reconfirming a uniform decay time constant (based on D0-$C_{CaK}$).

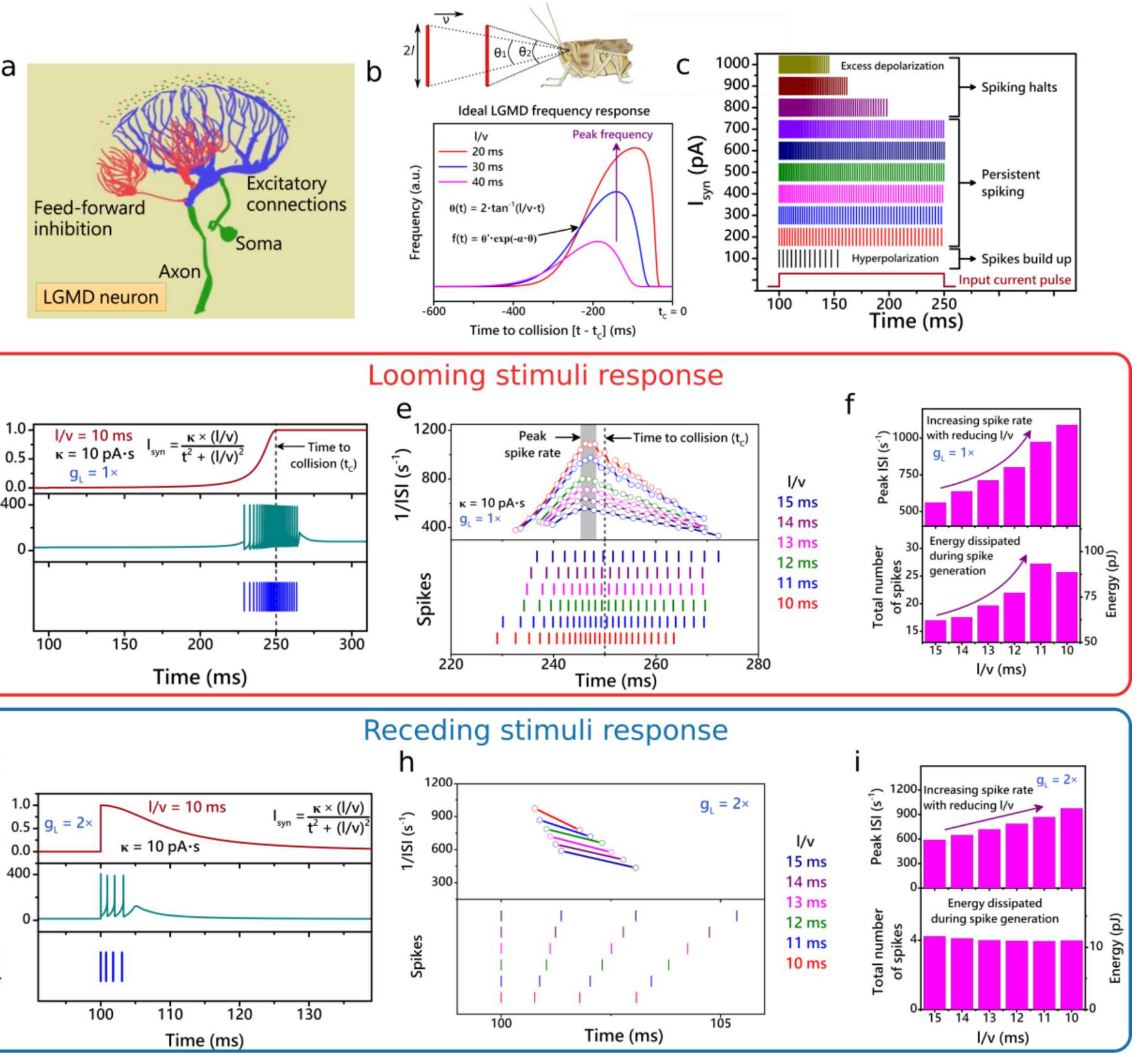

**Figure 4. Obstacle tracking and collision detection with LGMD-like spike response. (a)** Illustration showing the vision system of a locust with the Lobula giant movement detector (LGMD) neuron. **(b)** LGMD neuron performs a complex computation with θ and θ' to stimulate an escape response based on the size (*l*) and speed (*v*) of an object. Firing pattern of an LGMD neuron can be approximated by the expression f(t) = θ'exp(-αθ). A smaller *l/v* represents a higher threat and elicits a larger response. **(c)** Firing pattern of the neuron with SFA and PIR shown in Figure 3a is plotted for a range (100-1000 pA) of $I_{syn}$ amplitudes. A low value of $I_{syn}$ of 100 pA resulted in discontinuation of spiking due to excess hyperpolarization by $I_{CaK}$. Sustained spiking was observed with moderate (200-700 pA) $I_{syn}$ values. Higher (800-1000 pA) $I_{syn}$ resulted in halting of spikes because of excess depolarization of the membrane. These traits, along with PIR, help the artificial neuron mimic LGMD behavior closely. **(d)** Response ($V_M$ and raster plot) of the circuit for an $I_{syn}$ pulse (100-250 ms time span) to mimic synaptic current into an LGMD neuron for a looming stimulus. Time to collision ($t_C$) is set at 250 ms after which the synaptic current remains constant. **(e)** Instantaneous spike rate (top) and spike time (bottom) for $I_{syn}$ inputs similar to (d) for a range of *l/v* values. Peak spike rate is achieved in each case prior to $t_C$. **(f)** Peak $ISI^{-1}$ and total number of spikes (and hence, energy) increase with decreasing l/v indicating that an input with higher threat triggers an enhanced response at the cost of higher energy consumption. **(g-i)** Similar data as in (d-f) for receding stimuli, and with 2× leakage conductance ($g_L$).

## Mimicking Lobula giant movement detector (LGMD) response

Many insects and even large animals contain specialized neurons in their visual neural pathway for obstacle detection. These neurons have the capability to detect objects moving towards them and trigger large firing rates when the potential for collision is detected based on the speed and direction of the incoming object. Such computations are extremely energy-, area- and time-sensitive in neurobiology. Small insects like locusts have developed compact and efficient collision detection neurons such as the LGMD. A single LGMD can detect collision of an incoming object within a short time (usually a few milliseconds) with good energy efficiency based on complex computations involving size, location and speed dynamics of the object. **Figure 4**a shows an illustration of the LGMD neuron along with important dendritic trees and neurons. A large dendritic network including lateral and feed-forward inhibitions is an important part of LGMD operation. Another neuron, the descending contralateral movement detector (DCMD), is an equally important part of the system which receives input from the LGMD and can trigger motor neurons if a collision is detected. Neuromorphic algorithms based on LGMD neuron behavior have been implemented for obstacle detection and avoidance in mobile robots and automotive applications[48-51]. Figure 4b shows how LGMD computes a complex function (encoded as the spike rate) based on the dynamics of the angle an object makes on the visual field of the locust. Time axis is plotted with respect to time-to-collision ($t_C$). $\theta$ depends on the size ($l$) and speed ($v$) of the object as $\theta(t) = 2\tan^{-1}(l/v \cdot t)$. Firing rate of the LGMD neuron can be estimated by the expression $f(\theta) = \theta' \cdot \exp(-\alpha\theta)$. Here, the parameter $l/v$ determines the spike response such that a smaller $l/v$ (a small but fast-moving object) elicits a higher firing rate implying a larger threat.

Biological LGMD neuron demonstrates many features like SFA, PIR, after-hyper potential, initial burst response to a small input current, and halting of spikes[29]. It is important to reproduce these functionalities in an artificial neuron to mimic LGMD response closely with speed- and energy-efficiency. The circuit in Figure 3a was able to match all of the traits listed above, and hence, can be used to implement complex LGMD behavior with appropriate input current modeling. Figure 4c shows the circuit response to a wide range (100-1000 pA) of $I_{syn}$ magnitudes. For a small $I_{syn}$ value of 100 pA, a limited number of spikes were recorded before the spiking activity stops due to excess hyperpolarization caused by increased $I_{CaK}$. This closely mimics the behavior of LGMD at relatively small input excitation. On the other hand, the spiking is sustained for the entire duration of the current pulse for intermediate (200-700 pA) $I_{syn}$ values with SFA. Finally, for large (800-1000 pA) $I_{syn}$ values, spiking halts because of excess depolarization, which can help to achieve a drop in the firing rate based on the synaptic current levels when approaching $t_C$.

Synaptic current generated by the input dendritic network of the LGMD neuron for a looming (approaching) object can be modelled based on the angular size (θ) of the looming object as $I_{syn} \propto \theta'$, which can be expressed in terms of the parameter $l/v$ as $I_{syn}(t) = \frac{\kappa\cdot(l/v)}{t^2+(l/v)^2}$, where κ is a proportionality constant[39]. Figure 4d shows $V_M$ response (middle panel) of the neuronal circuit and a raster plot (bottom panel) for the $I_{syn}(t)$ waveform (top panel) with $\kappa = 10$ pA·s, $l/v = 10$ ms, and $t_C = 250$ ms. The input current is applied after an initial delay of 100 ms in each simulation to enable the circuit to relax to the rest state. Mimicking the biological LGMD, the artificial neuron issued spikes once the input current (i.e., threat of the looming stimulus) was sufficiently large. Spike rate ($ISI^{-1}$) peaked before $I_{syn}$ reached its maximum value near $t = t_C$ and then dropped. The plots on the right show the same data on a zoomed-in time axis (210-290 ms) for better clarity. The same exercise with $t_C = 250$ ms was repeated for multiple $l/v$ values (10-15 ms) and the spike response is shown in Figure 4e as a raster plot (bottom panel) along with the corresponding instantaneous $ISI^{-1}$ values plotted in the top panel. Peak spike rate increases monotonically with decreasing *l/v* (increasing threat) as observed in biological LGMD. Furthermore, the peak $ISI^{-1}$ value is reached before $t_C$, suggesting successful detection of collision in each case.

Peak $ISI^{-1}$ generated for each of the cases in Figure 4e is plotted in the top panel of Figure 4f. The total number of spikes and corresponding energy dissipation during the entire duration of the $I_{syn}$ signal are plotted in the bottom panel. Increasing number of spikes along with increasing spike rate for smaller *l/v* values represent an elevated collision threat and uses higher energy. The total amount of energy dissipated by the circuit for the duration of over 300 ms does not exceed ~100 pJ for the smallest *l/v* value. This observation indicates how computationally complex and time-sensitive tasks can be performed using bio-inspired neuromorphic solutions with minimal energy cost. The recorded energy consumption of ~100 pJ with LGMD spike response improves both the power consumption and ease-of-integration in spiking neural networks over contemporary reports on mimicking LGMD operation[30,31]. The response of the circuit for receding objects (similar to Figures 4d-f) is shown in the SM Figure S9. Spike response to $I_{syn}$ mimicking receding objects showed smaller spike count along with a shorter duration of response. These traits are also observed in biological LGMD neurons enabling them to differentiate between looming and receding objects[28]. As mentioned earlier, inherent selection of collision-sensitive stimuli adds another layer of safety and can ensure quick response time with low energy dissipation. Additionally, the spike behavior for $I_{syn}$ corresponding to a looming stimulus can be further tuned (for the same *l/v* values) by appropriate choice of the sensor proportionality constant κ as shown in the SM Figure S10. Such tuning can enable the use of the circuit in applications with limitations on energy consumption (small spike rate with small energy dissipation) or applications requiring time-critical response (large spike rate with higher energy dissipation) without any physical change in the circuit of the neuron. Spike response from the artificial neuron circuit can be further modulated by choosing an appropriate conductance $g_L$ of the leakage transistor $T_L$. SM Figure S11 shows how spike count and peak $ISI^{-1}$ shift from a looming-selective response towards a response only for receding stimuli (zero spikes for looming stimuli) as the leakage conductance is increased. Selectivity modulation via circuit design can enable a neuron network that can provide complete information of the speed and type of movement (looming *vs* receding). Figures 4g-i show circuit response to receding stimuli with monotonically increasing peak $ISI^{-1}$ as the *l/v* value decreases with a constant spike count, and hence the dissipated energy of ~11 pJ. Instantaneous and peak $ISI^{-1}$ values along with the spike count in the output can enable accurate detection and tracking of the looming as well as receding movements of objects with high energy efficiency.

# Conclusion

In summary, novel subthreshold operation of electrostatically controlled 2D channel transistors and its application in biomimetic artificial neuron design have been demonstrated. The device operation is shown to be unaffected by the choice of 2D channel material and its dominant transport polarity (n- or p-type). The 2D-ST can be electrostatically controlled by two gate-biases ($V_{GC}$, $V_{GB}$) to give a tunable, low-current bell-shaped subthreshold current that mimics the Na-channel current observed in biological neurons. Device current from the subthreshold operation of 2D-ST has been modeled and fitted using an over-the-barrier thermionic current controlled by $V_{GC}$, and a Fowler-Nordheim tunneling current controlled by $V_{GB}$. These physics-based fits have been further corroborated with calculations based on the geometrical dimensions of the 2D-ST. Simulation results of a circuit designed with 45 nm pdk components and the 2D-ST exhibit neuron operation closely matching the Hodgkin-Huxley model for biological neurons. Energy per spike as low as ~3.5 pJ was observed with monotonically increasing spike rate for class-I spike behavior. The same circuit can also be used as an oscillatory neuron after appropriately adjusting the leakage current component. Additionally, the circuit was modified to demonstrate spike frequency adaptation and post-inhibitory rebound behaviors by mimicking the $Ca^{2+}$-activated K-channel polarizing current that controls both the spike rate as well as the membrane potential relaxation time scale. SFA and PIR are important functionalities observed in collision detecting biological LGMD neurons. LGMD in insects like locusts is sensitive to looming (approaching) objects, and can trigger spike rate-coded firing patterns to indicate possible collision threat based on the size and speed of the object. The artificial neuron circuit with the 2D-ST is shown to achieve all important functionalities of LGMD neurons. The simulation results demonstrate that the artificial neuron is able to detect approaching objects well ahead of collision with limited number of spikes and total energy consumption less than ~100 pJ. The same circuit can also differentiate between looming and receding stimuli, much like a biological LGMD neuron. Furthermore, the ability to tune the selectivity of the LGMD neuron circuit response to looming *vs* receding objects adds another degree of freedom for design of real-time multi-object tracking systems. This work can potentially be used for low-power obstacle tracking and avoidance in autonomous vehicles using spiking neural networks. Additionally, it is useful for implementing energy-efficient spiking or oscillatory neuron networks as well as dynamic sensory and memory system applications involving SFA or PIR functionalities.

## Methods

**Device fabrication.** The devices were fabricated on 4 in. Si/(285 nm)$SiO_2$ substrate wafers. First, the channel gate ($G_C$) and the barrier gate ($G_B$) were patterned using electron beam lithography (EBL) in Raith 150-Two. The two $G_B$ electrodes were shorted during the EBL by design. Cr/Au (3/30 nm) metallization for gate electrodes was done in an ATC Orion (a 7-target sputtering system) by AJA International followed by a lift-off process. hBN and 2D channel material ($MoS_2$, $WSe_2$, $WS_2$) flakes were mechanically exfoliated onto scotch tape from crystals purchased from SPI Supplies. Afterwards, these hBN and 2D channel material flakes were transferred onto the substrate using polydimethylsiloxane (PDMS) assisted dry transfer method and locally aligned on the structure under a microscope objective. Next, the source/drain (S/D) electrodes were patterned using EBL such that there is an overlap with the $G_B$s on both sides. This allows for better control over contact barriers by the $G_B$. Lastly, Cr/Au (3/100 nm) was sputtered for the devices with n-type $MoS_2$ and $WS_2$; and Cr/Pt/Au (3/30/100 nm) was sputtered on the device with p-type $WSe_2$ for the S/D contacts.

**Device characterization.** AFM measurements for $MoS_2$ and hBN flakes were carried out in an MFP-3D system by Oxford Instruments. All steady-state electrical data was recorded under ambient conditions using a B1500A semiconductor device parameter analyzer module by Keysight. Bell-shaped curves were recorded by simultaneously sweeping two voltage channels in the B1500A.

**Circuit simulations and analysis.** All circuit simulations were performed using Spectre (Cadence). The scaled data from the 2D-ST measurements was input as a Verilog-A model. All transistors and MOSCAPs were taken from a general purpose 45 nm process design kit (gpdk045). The results were collected as comma-separated values (csv) files and processed in MATLAB for subsequent analysis and parameter extraction.

## Acknowledgements

SL acknowledges funding support from the Department of Science and Technology (DST), Govt. of India (DST/SJF/ETA-01/2016-17). BR acknowledges funding support from Engineering and Physical Sciences Research Council (EPSRC), United Kingdom (EP/X011356/1).

## Author contributions

KT proposed the device idea and carried out fabrication and characterization of the devices. KT, BR and SL discussed and analyzed the device data. KT, BR and SL prepared the circuit simulation plan. KT performed simulations and analyzed the data along with BR and SL. All authors discussed the results and wrote the manuscript. SL and BR directed the research.

## Supplementary information

AFM scans of the 2D flakes, physics-based fitting of the device characteristics under subthreshold regime, electrical characteristics of the $WS_2$ channel transistor, Hodgkin-Huxley model equivalent circuit response, leaky integrate-and-fire response of the artificial neuron, artificial neuron behavior without the bias scaling of the $WSe_2$ transistor, details of the component specifications used for the circuit simulations, analysis of the decay of the post-inhibitory response, LGMD response to the receding stimuli, tuning of response to the looming and receding stimuli, tuning of LGMD response with leakage path conductance.

Supplementary Material for

# Ultra-Low Power Neuromorphic Obstacle Detection Using a Two-Dimensional Materials-Based Subthreshold Transistor

Kartikey Thakar[1], Bipin Rajendran[2], and Saurabh Lodha[1,*]
[1]Department of Electrical Engineering, Indian Institute of Technology Bombay, Mumbai, India.
[2]Department of Engineering, King's College London, Strand, London, United Kingdom.
*slodha@ee.iitb.ac.in

## Contents

## Section S1. AFM data of $MoS_2$, $WSe_2$ and hBN flakes

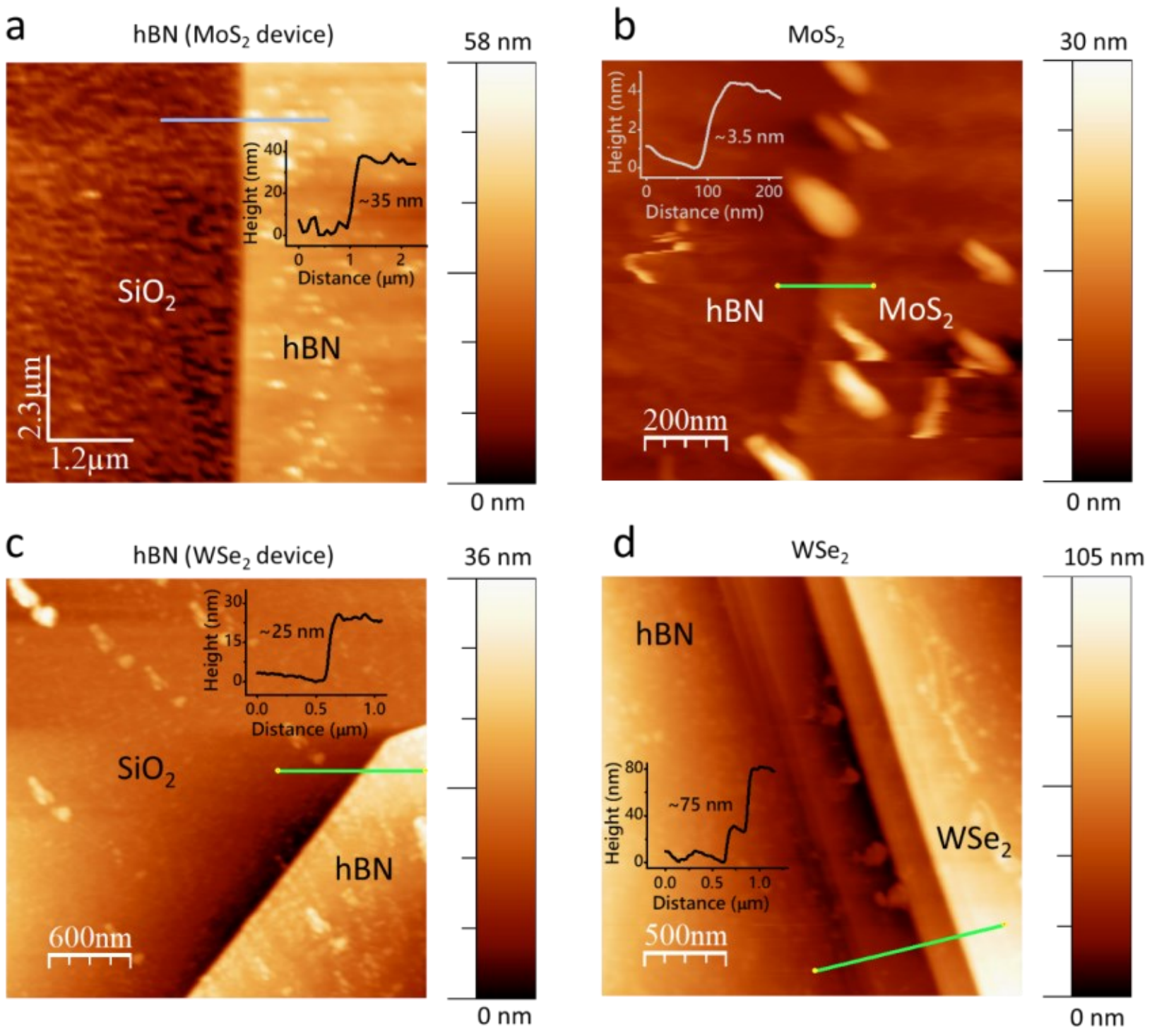

**Figure S1.** AFM images along with the line scans showing thickness of the **(a)** hBN ($MoS_2$ device), **(b)** $MoS_2$, **(c)** hBN ($WSe_2$ device), and **(d)** $WSe_2$ flakes.

## Section S2. Current-voltage map for the 2D-ST device with $WS_2$ channel

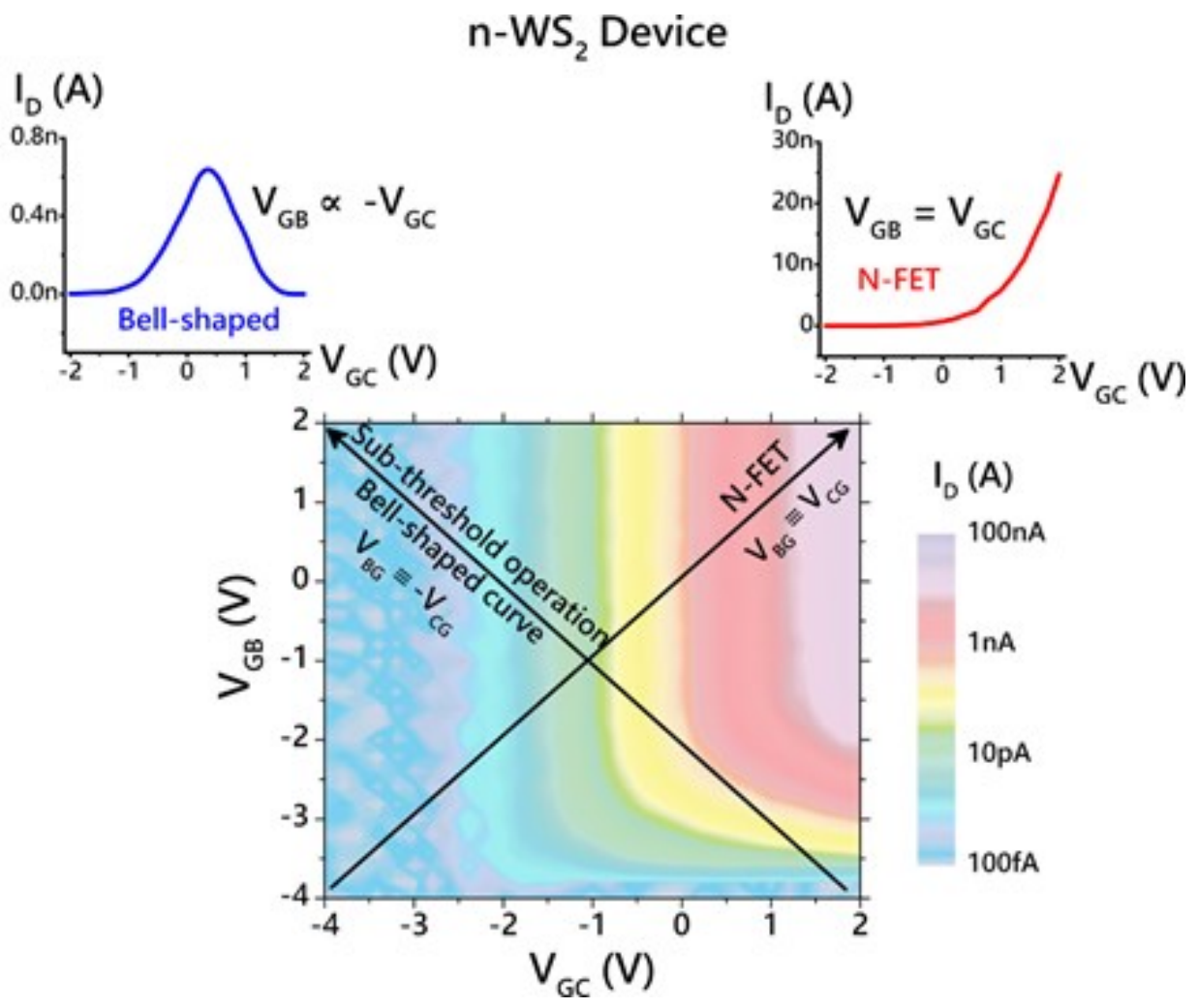

**Figure S2.** Colormap showing device current of an n-$WS_2$ 2D-ST device as a function of the channel and barrier gate biases ($V_{GC}$, $V_{GB}$). The line scan along $V_{GC} = V_{GB}$ direction from the current map can achieve a conventional n-FET transfer curve. Whereas, taking a line scan along the opposite direction ($V_{GC} \equiv -V_{GB}$) gives a bell-shaped curve under subthreshold operation.

## Section S3. Physics-based fitting of the bell-shaped I-V curves

Bell-shaped I-V curves were obtained by simultaneously sweeping the two gate-bias ($V_{GC}$, $V_{GB}$) values in opposite directions. The device operates in subthreshold regime during the entire I-V sweep. Device equivalent circuit and band diagrams are shown in Figure S3-1a based on a lumped resistor model representing subthreshold thermionic current (controlled by $V_{GC}$) and Fowler-Nordheim (FN) tunneling current (controlled by $V_{GB}$) in series. The operating regimes where each of these resistors ($R_{FN}$ and $R_{ST}$) limit the device current are marked on the I-V curve in Figure S3-1b.

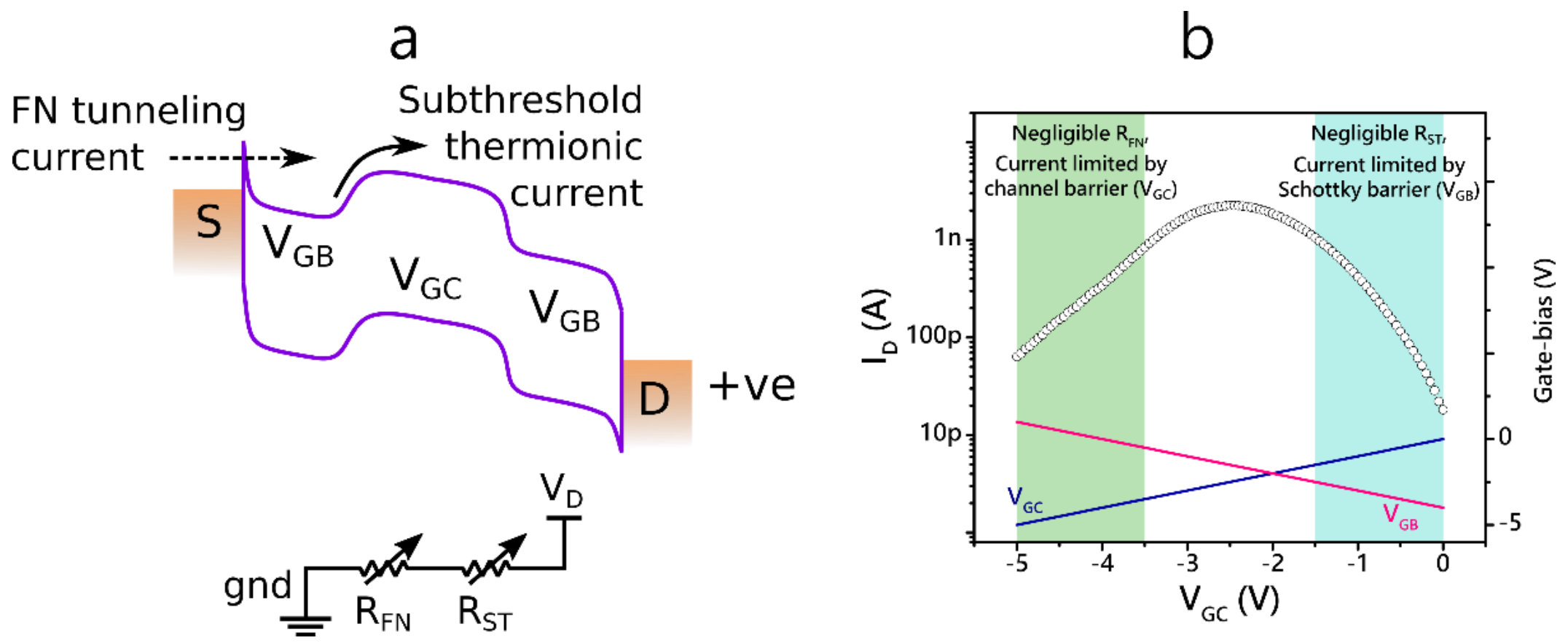

**Figure S3-1. (a)** Band diagram of the 2D channel showing the two main current components that determine the current flowing through the device. The bottom schematic is an equivalent lumped resistor model for the device. **(b)** Bell-shaped subthreshold current behavior for given gate-bias ($V_{GC}$, $V_{GB}$) signals from the $MoS_2$ device. The two regimes where current is limited by either $V_{GC}$ or $V_{GB}$ are marked.

Contact resistance estimated using the Y-function method is approximately $10^6$ Ω corresponding to a current of ~1 μA at $V_D$ = 1V. Hence, while modelling the bell-shaped I-V with peak current levels <10 nA, we can ignore the inclusion of a lumped $R_c$ value.

The two variable resistors can be modelled by equations governing the respective transport mechanisms as discussed below.

**(a) Subthreshold thermionic current** (Figure S3-2a):

$R_{ST}$ is dominant on the left side of the bell-shaped curve where $V_{GB}$ is large (small tunneling barrier) and $V_{GC}$ is small (large thermionic barrier). Hence, the device current can be modelled using the following equation:

$$I_{ST} = A \cdot (m-1) \cdot \exp\left( V_{GC} - \frac{V_{Th}}{m \cdot \left(\frac{kT}{q}\right)} \right)$$

Here, $A$ (current proportionality constant), $m$ (body-factor) and $V_{Th}$ (threshold voltage) are the fitting parameters.

**(b) FN tunneling current** (Figure S3-2b):

$R_{FN}$ is dominant on the right side of the bell-shaped curve where $V_{GC}$ is large (small thermionic barrier) and $V_{GB}$ is small (large tunneling barrier). Hence, the device current can be modelled by the following set of equations:

$$I_{FN} = \frac{A}{\phi}\exp\left(-B\cdot\phi^{\frac{3}{2}}\right), \text{ with}$$

$$\phi = \phi_0 - \Delta\phi;\text{ where } \Delta\phi = \sqrt{\frac{qE_m}{4\pi\epsilon}},\text{ and } E_m = \sqrt{\frac{2qN_D}{\epsilon}(\psi - V_A)}.$$

Substituting $\psi = V_{GB}/m$, we can write

$$\Delta\phi = a\cdot(V_{GB} - b)^{\frac{1}{4}}.$$

Here, $A$, $B$, $\phi_0$, $a$, and $b$ are the fitting parameters.

Results of the region-wise partial fits are shown in Figure S3-2. Based on the series resistance model, total device current can be then approximated as

$$I = \frac{1}{\frac{1}{I_{ST}} + \frac{1}{I_{FN}}}.$$

Fitting of the bell-shaped I-V curves for the entire gate-bias range based on the above expressions is shown in Figure S3-3.

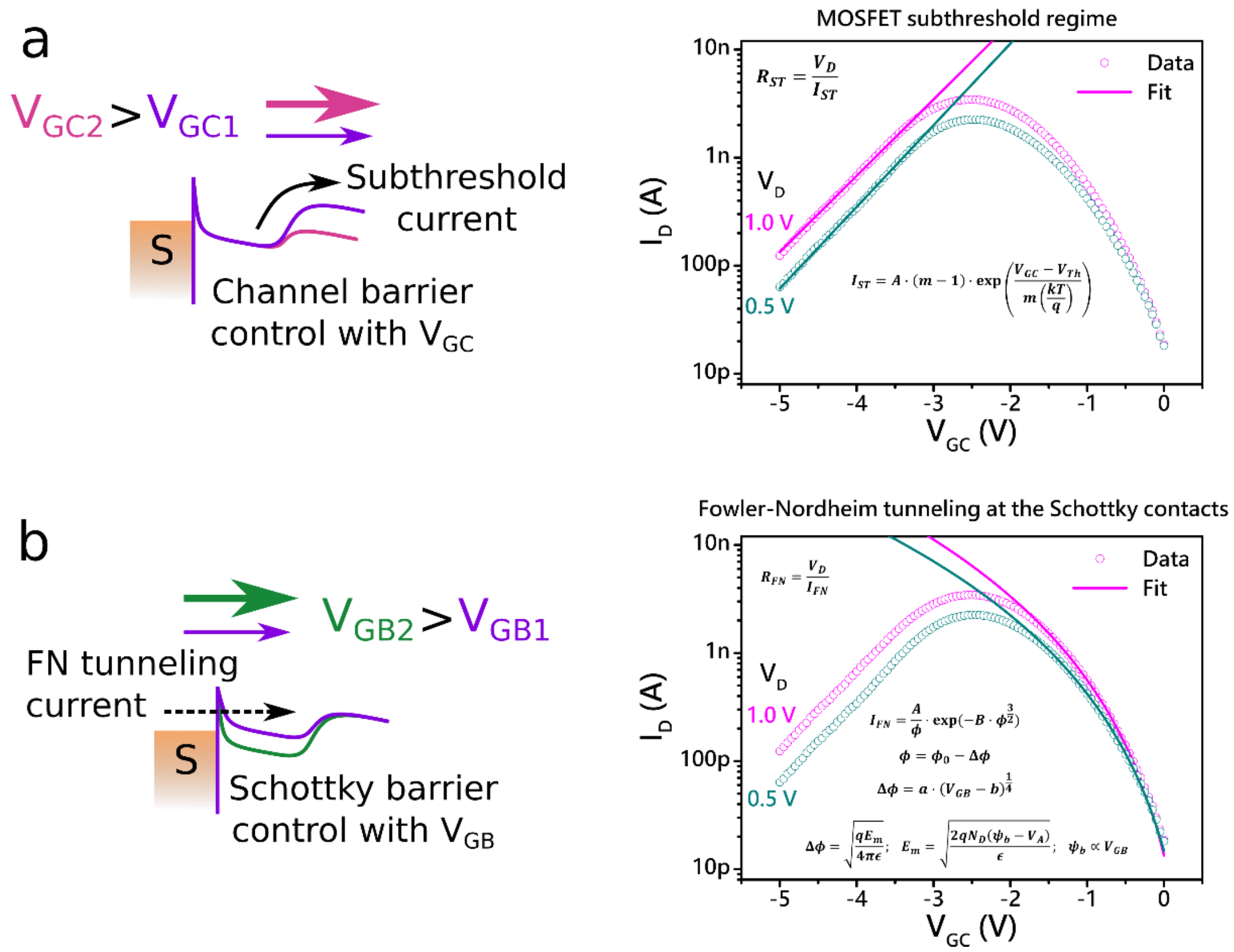

**Figure S3-2. (a)** Band diagram showing change in over-the-barrier thermionic current with $V_{GC}$ and partial fitting of the I-V curve using the exponential relation in the subthreshold regime. **(b)** Band diagram showing change in FN tunnelling current with $V_{GB}$ and partial fitting of the I-V curve using the FN tunnelling equations.

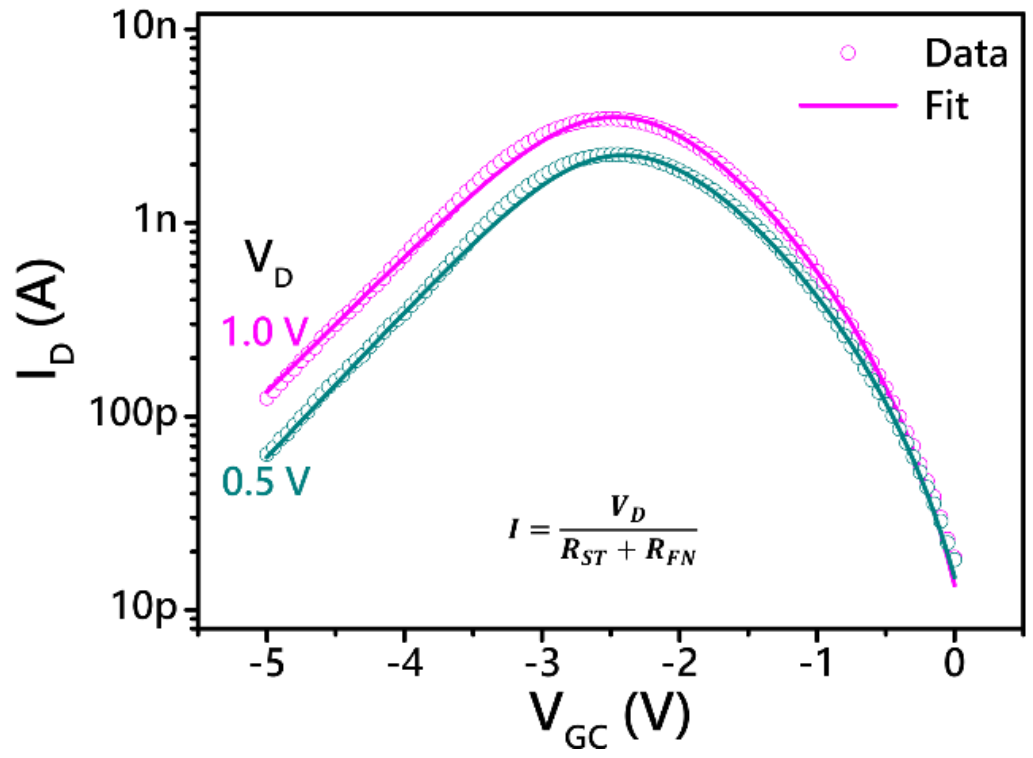

**Figure S3-3.** Fitting of the I-V curves for the full range of gate-bias.

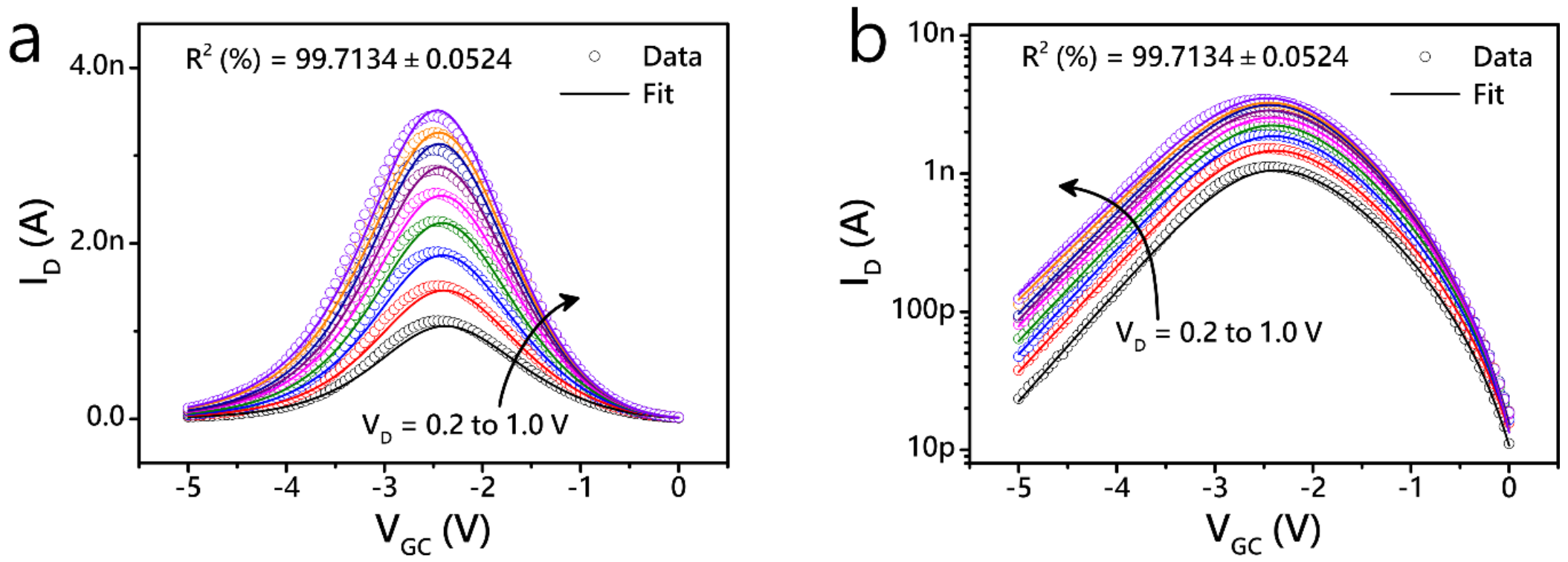

**Figure S3-4.** Fitting of the bell-shaped I-V curves for a range of $V_D$ values in **(a)** linear, and **(b)** log scale.

I-V curves for a large range of $V_D$ values were fitted (Figure S3-4) with the above-mentioned equations and the extracted important device parameters are listed below.

$$m = 22.52 \pm 0.95$$

$$\phi_0 = 0.164 \pm 0.002\ eV$$

Now, based on the flake thicknesses measured from AFM measurements of the hBN (~35 nm) and $MoS_2$ (~3.5 nm) flakes and taking the dielectric constant values of the two materials as $\epsilon_{hBN} = 3.8$[1], and $\epsilon_{MoS2} = 8$[2], we can estimate

$$m = 1 + \frac{C_D}{C_{ox}} = 1 + \frac{\epsilon_{MoS2} \cdot t_{hBN}}{\epsilon_{hBN} \cdot t_{MoS2}}$$

$$\Rightarrow m = 1 + \frac{8 * 35}{3.8 * 3.5} \approx 22,$$

which closely matches the extracted value of $m = 22.52 \pm 0.95$.

Next, by substituting the extracted value of $m$ = 22.52, and $t_{MoS2}$ = 3.5 nm (from AFM measurements) in the equation for $\Delta\phi$ with $a = 0.09542 \pm 0.00259$, we can write

$$a = \sqrt{\frac{q}{4\pi\epsilon_{MoS2}} \sqrt{\frac{2qn_{2D}}{m \cdot \epsilon_{MoS2} \cdot t_{MoS2}}}}, where\ N_D = \frac{n_{2D}}{t_{MoS2}}.$$

$$\Rightarrow n_{2D} \sim 4.5 \times 10^{12}\ cm^{-2}.$$

This extracted value of $n_{2D}$ for $MoS_2$ is in agreement with literature[3].

## Section S4. Hodgkin-Huxley (HH) model equivalent circuit response

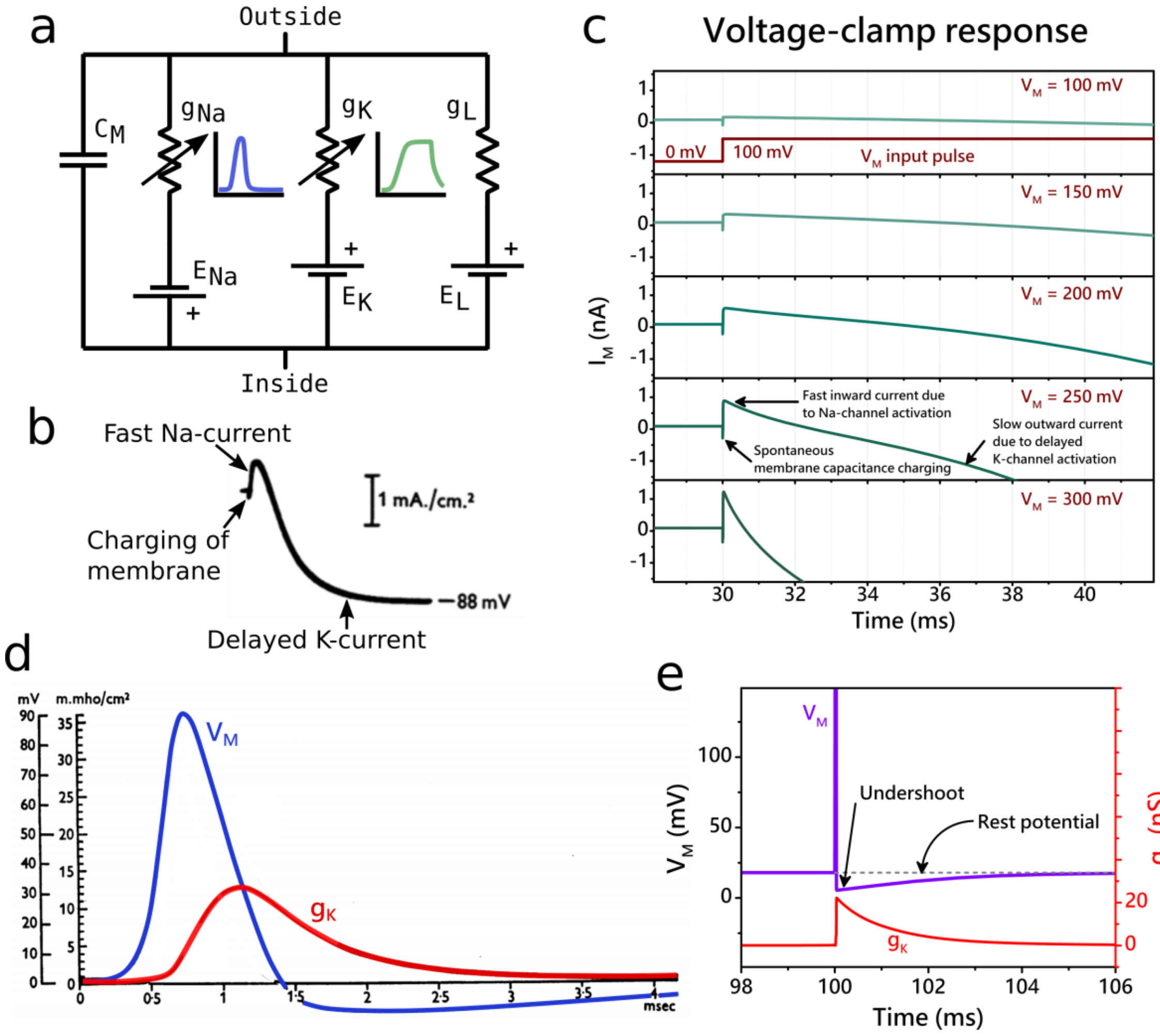

**Figure S4. (a)** Electrical equivalent circuit for the Hodgkin-Huxley model for biological neurons with a membrane capacitance and Na, K and leakage channels. **(b)** Membrane current measured in a biological neuron with membrane voltage ($V_M$) clamped at a fixed value. Fast inward (here shown as positive) current is due to Na-channel activation and slow outward (here shown as negative) is because of delayed K-channel activation. Both the magnitude and time scale depend on $V_M$[4]. **(c)** Voltage-clamp equivalent results from the circuit simulations showing $V_M$-dependent magnitude and time scale of the various current components. This shows similarity of the simulated circuit with the biological neuron model. **(d)** Membrane potential and potassium conductance ($g_K$) during a spike indicates that the membrane potential returns to its rest value after $g_K$ becomes zero[5]. **(e)** Circuit simulations for a single spike show that $V_M$ returns to its rest position after an undershoot when $g_K$ relaxes to zero.

## Section S5. LIF behavior of the artificial neuron

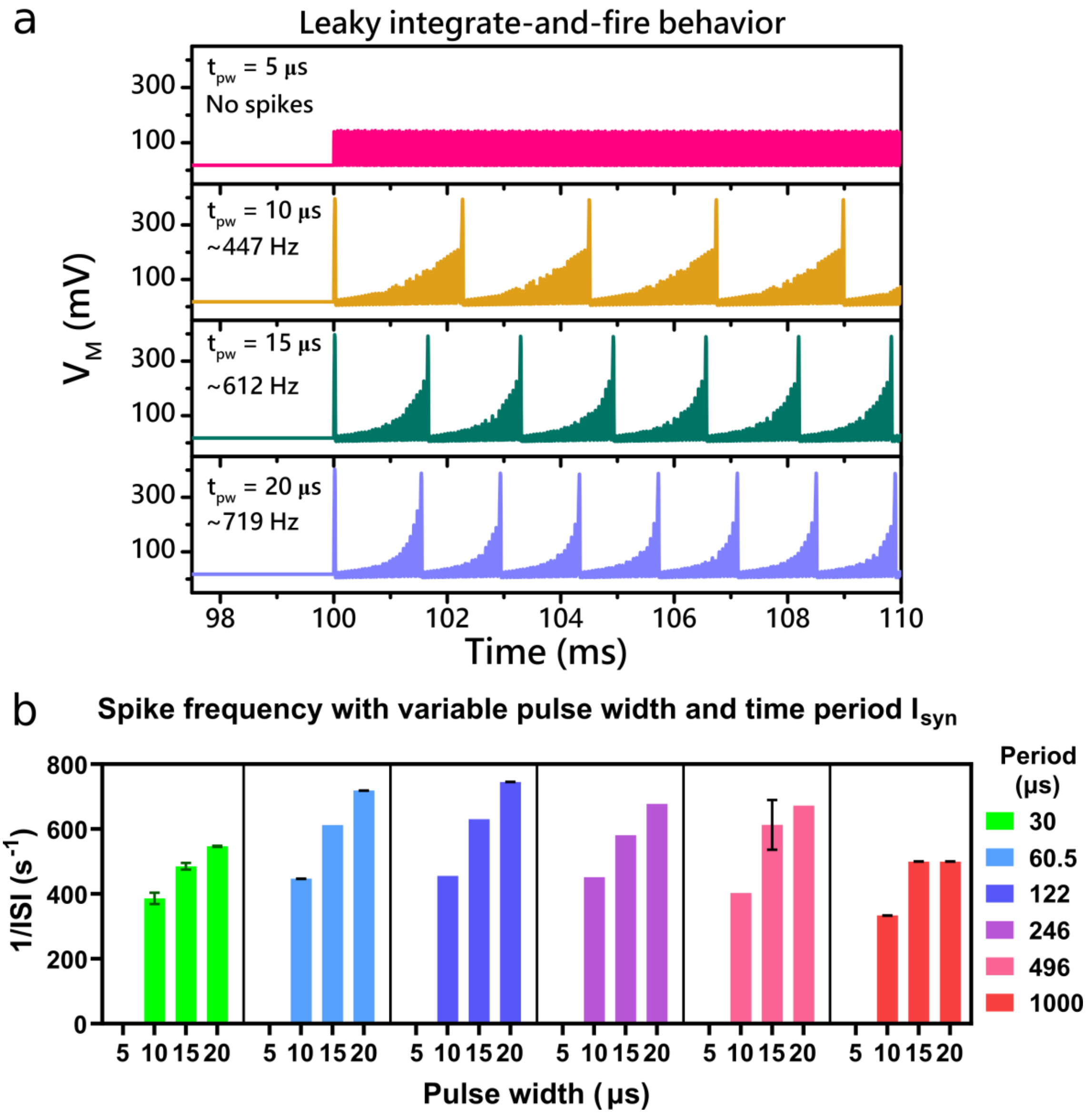

**Figure S5.** **(a)** Leaky integrate-and-fire neuron behavior showing increase in the spike rate with increasing pulse width of the $I_{syn}$ pulses with a time period of 60.5 µs. **(b)** Summary of the spike rate ($ISI^{-1}$) for $I_{syn}$ pulses with variable pulse width and time period. Increasing the separation between two consecutive pulses results in a decrease in the spike rate.

## Section S6. LIF neuron (RC) without gate-bias scaling of the $WSe_2$ transistor

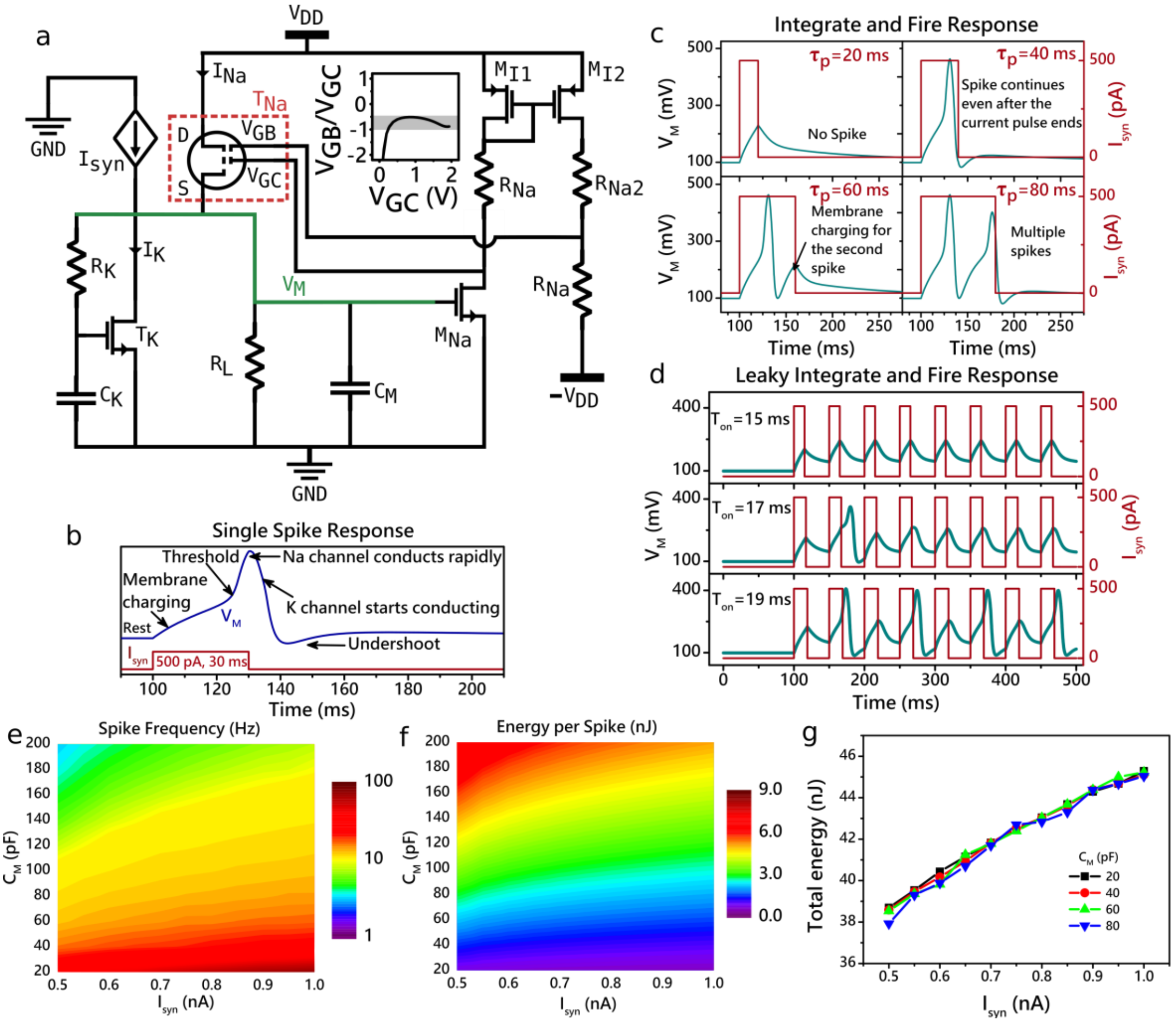

**Figure S6. (a)** Schematic of the neuronal circuit using the 2D-ST with p-$WSe_2$ ($T_{Na}$) providing biomimetic Na-channel current. Inset shows the relative variation in $V_{GB}$ with $V_{GC}$. Values for the circuit components are listed in SM Table S7-3. **(b)** Circuit simulation result for a 30 ms current pulse shows the generation of a single spike with observable markers of a biological neuron. Further circuit simulations show the circuit behavior as **(c)** an integrate-and-fire neuron for variable duration ($t_p$ = 20-80 ms) of current pulses, and **(d)** a leaky integrate-and-fire neuron with variable duty cycle ($T_{on}$ = 15-19 ms, $T_0$ = 50 ms) input pulses. Color maps showing evolution of **(e)** spike frequency, **(f)** energy per spike, and **(g)** total energy (over 1-second current pulse) as a function of input current magnitude and the capacitance values ($C_M$ = $C_K$). Energy per spike as low as ~500 pJ/spike was recorded with total power dissipation remaining unchanged with variation in $C_M$.

## Section S7. Details of component specifications used for the circuit simulations

**Table S7-1.** List of components for the circuit in Figure 2

<table>
<tr><th colspan="2">Component</th><th>Library/Model</th><th>Dimensions/Value</th><th>Unit</th></tr>
<tr><td colspan="2">$V_{DD}$</td><td>analoglib/vdc</td><td>1</td><td>V</td></tr>
<tr><td colspan="2">$I_{syn}$</td><td>analoglib/ipulse</td><td>0-1000</td><td>pA</td></tr>
<tr><td colspan="2">$MP1_{Na}$</td><td>gpdk045/pmos1v_hvt_3</td><td>120/45 (W/L)</td><td>nm</td></tr>
<tr><td colspan="2">$MP2_{Na}$</td><td>gpdk045/pmos1v_3</td><td>120/45 (W/L)</td><td>nm</td></tr>
<tr><td colspan="2">$MN1_{Na}$</td><td>gpdk045/nmos1v_hvt_3</td><td>240/45 (W/L)</td><td>nm</td></tr>
<tr><td colspan="2">$MN2_{Na}$</td><td>gpdk045/nmos1v_hvt_3</td><td>120/45 (W/L)</td><td>nm</td></tr>
<tr><td rowspan="4">$T_L$</td><td>$g_{L3}$</td><td rowspan="4">gpdk045/nmos1v_lvt_3</td><td>2400/45 (W/L)</td><td>nm</td></tr>
<tr><td>$g_{L2}$</td><td>2040/45 (W/L)</td><td>nm</td></tr>
<tr><td>$g_{L1}$</td><td>1800/45 (W/L)</td><td>nm</td></tr>
<tr><td>$g_{L0}$</td><td>1080/45 (W/L)</td><td>nm</td></tr>
<tr><td colspan="2">$T_K$</td><td>gpdk045/nmos1v_3</td><td>240/45 (W/L)</td><td>nm</td></tr>
<tr><td colspan="2">$MN_K$</td><td>gpdk045/nmos1v_3</td><td>120/45 (W/L)</td><td>nm</td></tr>
<tr><td colspan="2">$C_M$</td><td>gpdk045/nmoscap2v</td><td>50</td><td>fF</td></tr>
<tr><td colspan="2">$C_K$</td><td>gpdk045/nmoscap2v</td><td>200</td><td>fF</td></tr>
</table>

**Table S7-2.** List of components for the circuit in Figure 3

| Component | Library/Model | Dimensions/Value | Unit |
|---|---|---|---|
| $V_{DD}$ | analoglib/vdc | 1 | V |
| $I_{syn}$ | analoglib/ipulse | 0 to ±1000 | pA |
| $MP1_{Na}$ | gpdk045/pmos1v_hvt_3 | 120/45 (W/L) | nm |
| $MP2_{Na}$ | gpdk045/pmos1v_3 | 120/45 (W/L) | nm |
| $MN1_{Na}$ | gpdk045/nmos1v_hvt_3 | 240/45 (W/L) | nm |
| $MN2_{Na}$ | gpdk045/nmos1v_hvt_3 | 120/45 (W/L) | nm |
| $T_L$ | gpdk045/nmos1v_lvt_3 | 720/45 (W/L) | nm |
| $T_K$ | gpdk045/nmos1v_3 | 240/45 (W/L) | nm |
| $MN_K$ | gpdk045/nmos1v_3 | 120/45 (W/L) | nm |
| $C_M$ | gpdk045/nmoscap2v | 50 | fF |
| $C_K$ | gpdk045/nmoscap2v | 200 | fF |
| D0 | gpdk045/ndio2v | 200/200 (W/L) | nm |
| $C_{CaK}$ | gpdk045/nmoscap2v | 400 | fF |

**Table S7-3.** List of components for the circuit in SM Figure S6

| Component | Library/Model | Dimensions/Value | Unit |
|---|---|---|---|
| $V_{DD}$ | analoglib/vdc | 2 | V |
| $I_{syn}$ | analoglib/ipulse | 0-1000 | pA |
| $M_{Na}$ | gpdk045/nmos2v | 320/150 (W/L) | nm |
| $R_{Na}$ | analoglib/res | 125 M | Ω |
| $R_{Na2}$ | analoglib/res | 135 M | Ω |
| $M_{I1}$, $M_{I2}$ | gpdk045/pmos1v | 120/45 (W/L) | nm |
| $T_K$ | gpdk045/nmos2v | 320/150 (W/L) | nm |
| $R_K$ | analoglib/res | 300 M | Ω |
| $R_L$ | analoglib/res | 5 G | Ω |
| $C_M$ | analoglib/cap | 50 | pF |
| $C_K$ | analoglib/cap | 50 | pF |

## Section S8. Scaling of decay dynamics in PIR response

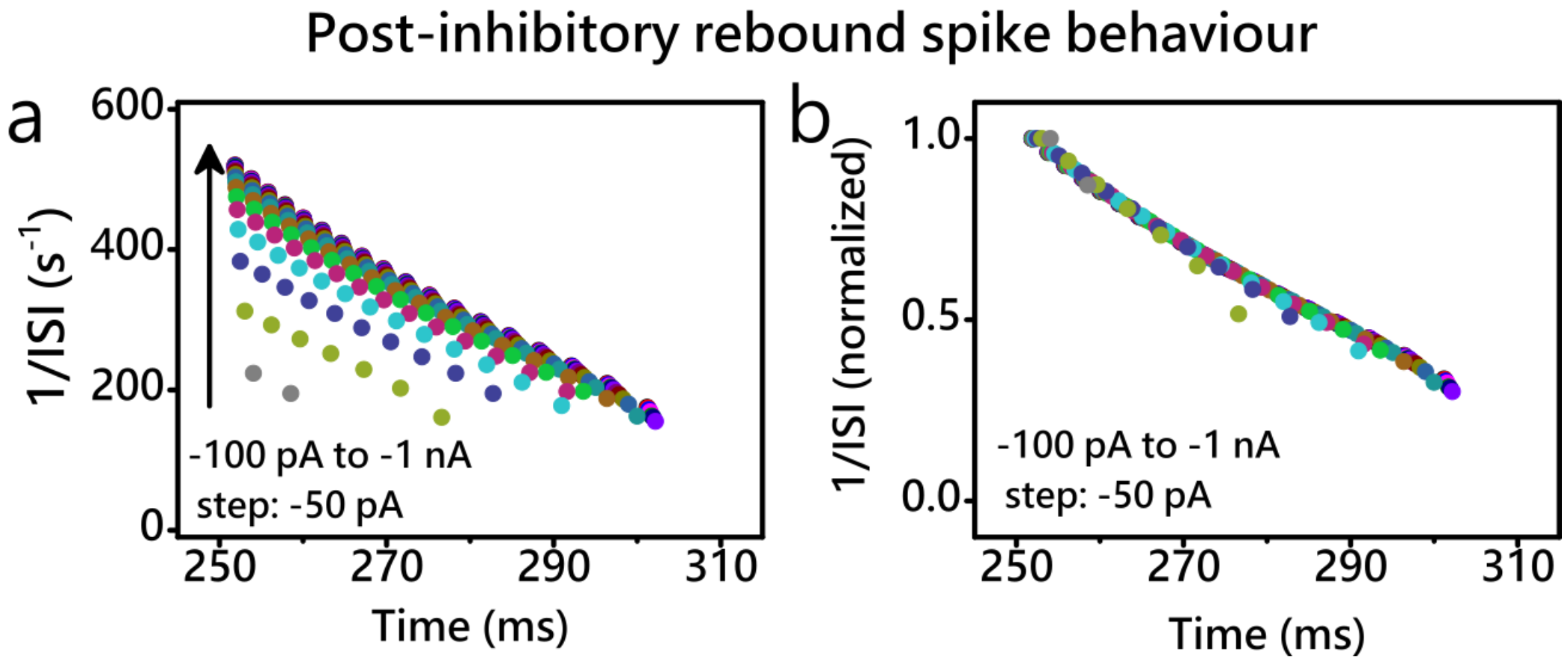

**Figure S8. (a)** Instantaneous spike rate as a function of time after inhibitory input shows decay within a short time. **(b)** Normalized spike rate for each of the $I_{syn}$ magnitudes reveals a single slope suggesting time or $I_{syn}$-independent uniform decay behavior.

## Section S9. Circuit response to receding stimuli

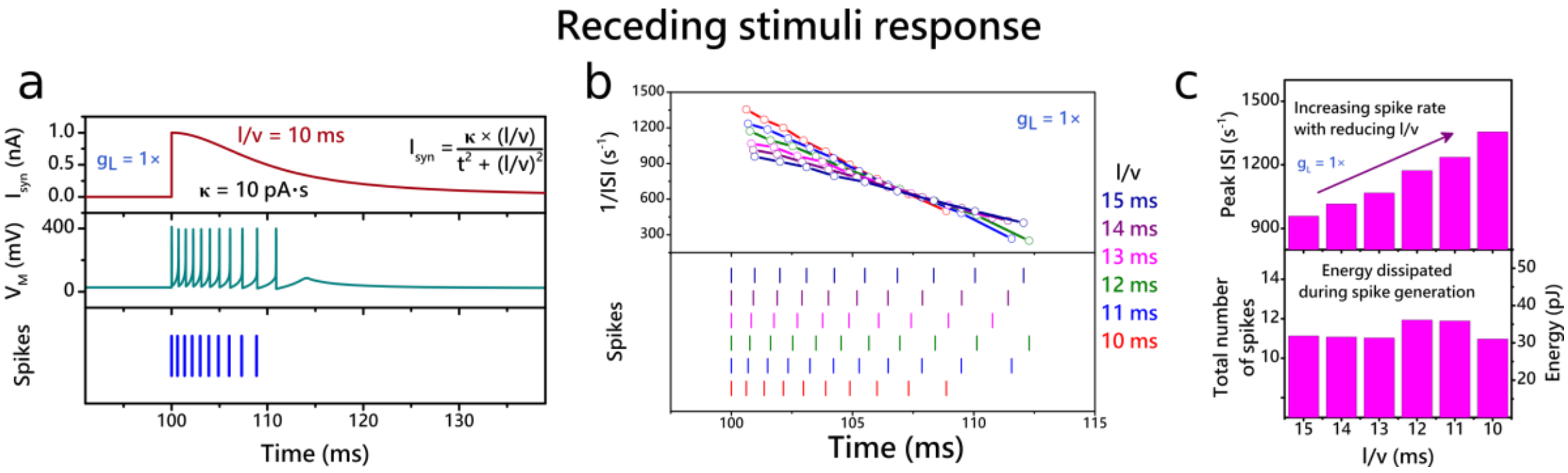

**Figure S9. (a)** Response ($V_M$ and raster plot) of the circuit for a $I_{syn}$ to mimic synaptic current into a LGMD neuron for a receding stimulus. **(b)** Instantaneous spike rate (top) and spike time (bottom) for $I_{syn}$ inputs similar to (a) for a range of *l/v* values. **(c)** Peak $ISI^{-1}$ increases with decreasing *l/v* suggesting that an input with large movement triggers an enhanced response. However total number of spikes (and hence, energy) does not change which helps LGMD response inherently differentiate between looming and receding response.

## Section S10. Tuning of spike response of looming and receding stimuli

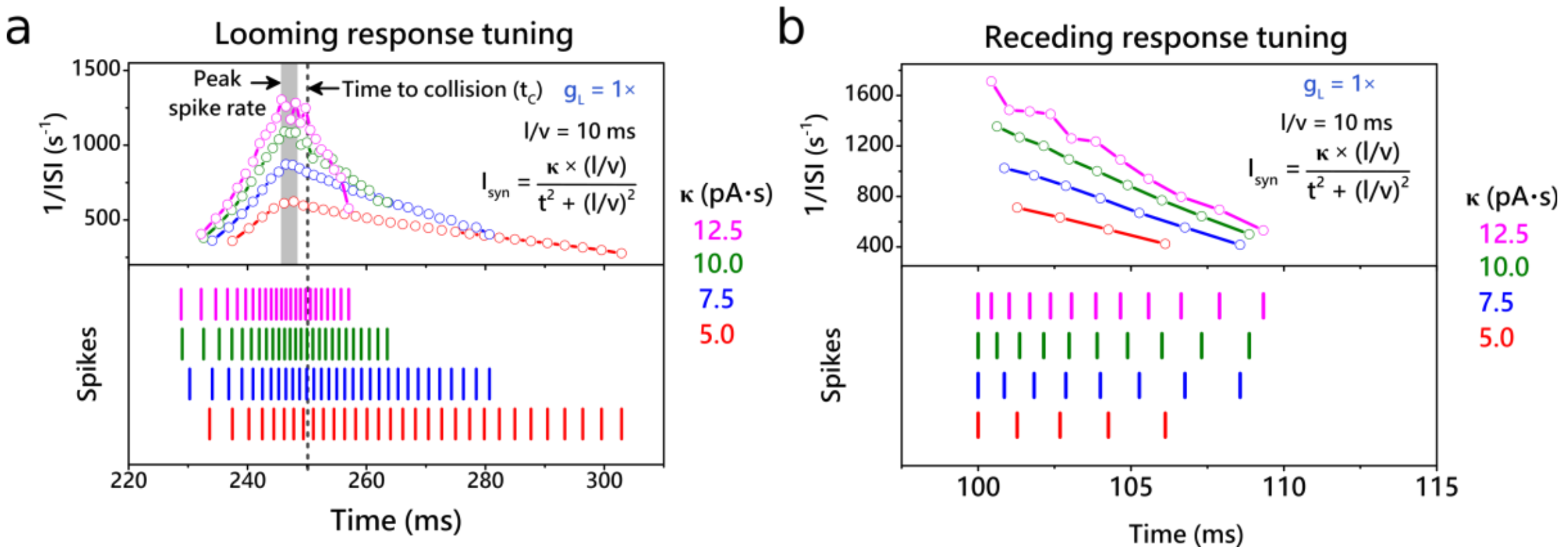

**Figure S10.** The frequency response for a receding stimulus can be tuned by setting an appropriate proportionality constant (κ) in the synaptic current to adjust the peak firing rate and duration of spiking activity. Instantaneous $ISI^{-1}$ and spike timing are shown for a range of κ values for both (a) looming, and (b) receding stimuli.

## Section S11. Selectivity of looming or receding stimulus

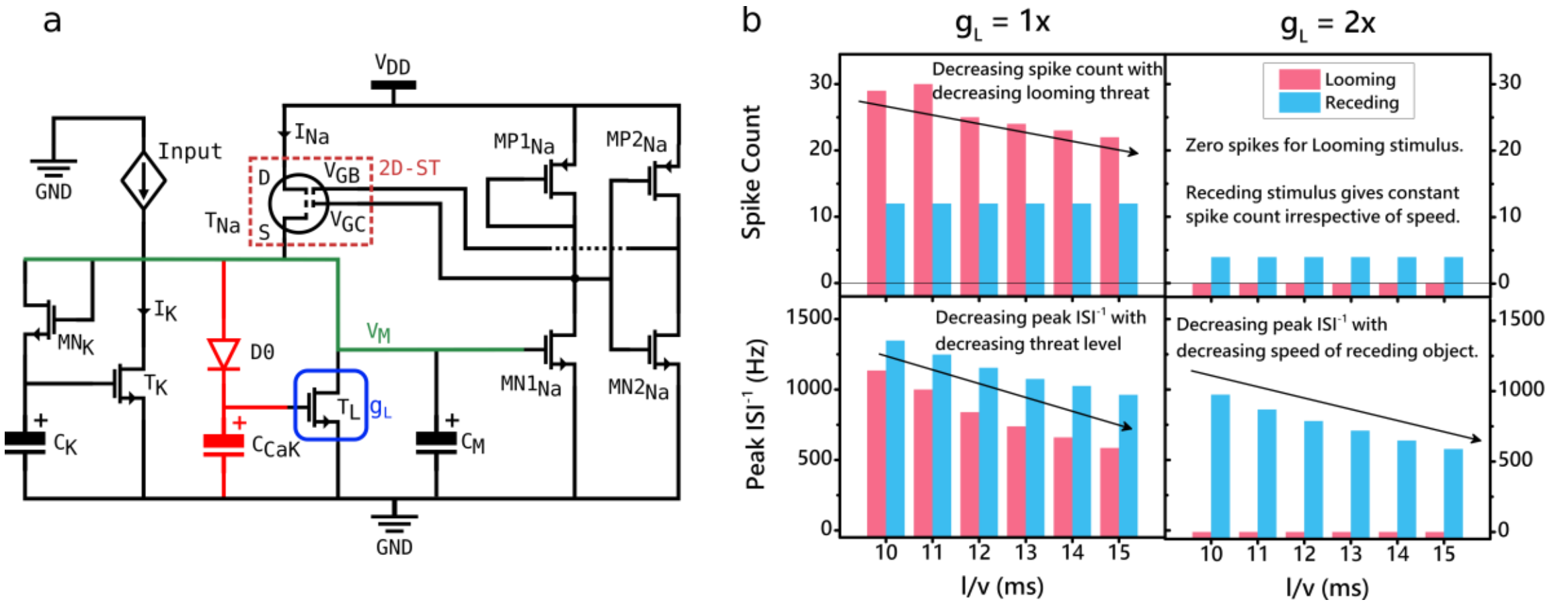

**Figure S11. (a)** Spiking neuron circuit same as shown in Fig 3a. Leakage path transistor $T_L$ with conductance $g_L$ is highlighted. **(b)** Spike count (top) and peak $ISI^{-1}$ (bottom) for $I_{syn}$ inputs for looming and receding stimuli ($l/v$ = 10 ms) for relative conductance values $g_L$ = 1× and $g_L$ = 2×. Both peak ISI and spike count decrease as the threat level decreases ($l/v$ increases). It should be noted that the spike count is zero for looming stimulus response for gL = 2×, making its response purely selective to receding stimulus. Such tuning via $g_L$ can be useful in employing direction selective (looming *vs* receding) object tracking.